\documentclass[twocolumn]{aastex63}
\usepackage{natbib}
\usepackage{multirow}
\usepackage{gensymb}
\usepackage{lipsum}
\usepackage{blindtext}
\usepackage{mathtools}
\usepackage{tensor}

\emergencystretch=\maxdimen
\shorttitle{Testing Rotating Regular Metrics with EHT Results of  Sgr A*}
\shortauthors{R.~K.~Walia et al.}
\definecolor{MyDarkBlue}{rgb}{0,0.08,0.5}
\definecolor{MyDarkRed}{rgb}{0.7,0.02,0.02}
\definecolor{MyDarkGreen}{rgb}{0.0,0.7,0.0}

\usepackage{amsmath}

\begin{document}
\title{Testing Rotating Regular Metrics with EHT Results of  Sgr A*}
	%
	
	\correspondingauthor{Rahul Kumar Walia}
	\email{rahul.phy3@gmail.com}
	\author[0000-0003-2471-1360]{Rahul Kumar Walia}
	\affiliation{ Astrophysics Research Centre, School of Mathematics, Statistics and Computer Science, University of KwaZulu-Natal, Private Bag 54001, Durban 4000, South Africa}
	\author[0000-0002-0835-3690]{Sushant G. Ghosh}
	\affiliation{Centre for Theoretical Physics, Jamia Millia Islamia, New Delhi 110025, India}
	\affiliation{ Astrophysics Research Centre, School of Mathematics, Statistics and Computer Science, University of KwaZulu-Natal, Private Bag 54001, Durban 4000, South Africa}
	\author[0000-0002-1967-2849]{Sunil D. Maharaj}
	\affiliation{ Astrophysics Research Centre, School of Mathematics, Statistics and Computer Science, University of KwaZulu-Natal, Private Bag 54001, Durban 4000, South Africa}

\begin{abstract}
The Event Horizon Telescope (EHT) observation unveiled the first image of supermassive black hole Sgr A* showing a shadow of diameter $\theta_{sh}= 48.7 \pm 7\,\mu$as with fractional deviation from the Schwarzschild black hole shadow diameter $\delta = -0.08^{+0.09}_{-0.09}~\text{(VLTI)},-0.04^{+0.09}_{-0.10}~\text{(Keck)}$. The Sgr A* shadow size is within $~10\%$ of the Kerr predictions, providing us another tool to investigate the nature of strong-field gravity. We use the Sgr A* shadow observables to constrain metrics of four independent and well-motivated, parametrically different from Kerr spacetime, rotating regular spacetimes and the corresponding no-horizon spacetimes. We present constraints on the deviation parameter $g$ of rotating regular black holes. The shadow angular diameter  $\theta_{sh}$ within the $1 \sigma$ region places bounds on the parameters $a$ and $g$. Together with EHT bounds on the $\theta_{sh}$ and $\delta$ of Sgr A*, our analysis concludes that the three rotating regular black holes, i.e., Bardeen  Hayward, and Simpson-Visser black holes, and corresponding no-horizon spacetimes agree with the EHT results of Sgr A*. Thus, these three rotating regular spacetimes and Kerr black holes are indiscernible in some parameter space, and one cannot rule out the possibility of the former being strong candidates for the astrophysical black holes.

\end{abstract}




\section{Introduction}
Black holes are one of the most profound predictions of the theory of general relativity (GR) \citep{Einstein:1915ca}, which were initially thought to be a mathematical consequence of GR rather than physically relevant objects \citep{Schwarzschild:1916uq}. However, later it was found that they are a generic and mostly inevitable outcome of gravitational collapse \citep{Oppenheimer:1939ue,Penrose:1964wq}. The no-hair theorem epitomizes the remarkable features of the GR black hole, stating that the Kerr (\citeyear{Kerr:1963ud}) and Kerr-Newman \citep{Newman:1965my} are the only stationary, axially symmetric, and asymptotically flat, respectively,  vacuum and electro-vacuum solutions to the Einstein equations  \citep{Israel:1967wq,Israel:1967za,Carter:1971zc,Hawking:1971vc,Robinson:1975bv}. 
Black holes embedded in the optically thin accreting region are expected to reveal a dark ``shadow"  encircled by a bright ring \citep{bardeen1973,CT}. The shadow's boundary is defined by the photon ring, which explicitly depends on the black hole parameters \citep{Johannsen:2010ru}. Therefore, the photon ring structure is an essential tool to test the theories of gravity through the astrophysical black holes' observational predictions.  Applications of shadow in understanding the near-horizon geometry have aroused a flurry of activities analyzing, both analytically and numerically, shadows for black holes in GR  \citep{Falcke:1999pj,Vries2000TheAS,Shen:2005cw,Yumoto:2012kz,Atamurotov:2013sca,Abdujabbarov:2015xqa,Cunha:2018acu,Kumar:2018ple,Afrin:2021ggx,Hioki:2009na} and modified theories  of gravity (MTGs) \citep{Amarilla:2010zq,Amarilla:2011fx,Amarilla:2013sj,Amir:2017slq,Singh:2017vfr,Mizuno:2018lxz,Allahyari:2019jqz,Papnoi:2014aaa,Kumar:2020hgm,Kumar:2020owy,Ghosh:2020spb,Afrin:2021wlj,Vagnozzi:2022moj,Vagnozzi:2019apd,Afrin:2021imp,Afrin:2021wlj}, and found the shadows of black holes in MTGs to be smaller and more distorted when compared with the Kerr black hole shadow. Also, the shadow of rotating regular spacetimes, which may be realized when GR is coupled to nonlinear electrodynamics, has been widely investigated to extract limits on the black hole magnetic charge \citep{Allahyari:2019jqz,Kumar:2020yem,Kumar:2020ltt,Kumar:2018ple,Amir:2016cen,Kumar:2019pjp}. In addition, by observing the size and deformation of the shadow, the spin, mass parameter, and possibly other global charges of the black holes can be estimated \citep{Hioki:2009na,Tsupko:2017rdo,Cunha:2019dwb,Cunha:2019ikd,Kumar:2018ple,Khodadi:2020jij,Ghosh:2020spb}. Besides, the black hole shadow is also valuable for testing theories of gravity. 

The event horizon is accessible only for tests with strong field phenomena near black holes, e.g., with the black hole shadow and gravitational waves. In 2019, the Event Horizon Telescope (EHT) collaboration released the first horizon-scale image of the M87* supermassive black holes  \citep{Akiyama:2019cqa, Akiyama:2019fyp,Akiyama:2019eap} which made black holes a physical reality. Using the M87* distance from the Earth $D=16.8$ Mpc and estimated mass  $M=(6.5 \pm 0.7) \times 10^9 M_\odot$,  bounds could be placed on the compact emission region size with angular diameter $\theta_d=42\pm 3\, \mu $as along with the central flux depression with a factor of $\gtrsim 10$, and circularity deviation $\Delta C\leq 0.10$. Recently, the EHT published the Sgr A* black hole shadow results, showing shadow angular diameter $\theta_{sh}= 48.7 \pm 7\,\mu$as with surrounding bright and thick emission ring of diameter $\theta_d=51.8\pm 2.3\mu$as \citep{EventHorizonTelescope:2022exc,EventHorizonTelescope:2022urf,EventHorizonTelescope:2022vjs,EventHorizonTelescope:2022wok,EventHorizonTelescope:2022xnr,EventHorizonTelescope:2022xqj}. Considering a black hole of mass $M = 4.0^{+1.1}_{-0.6} \times 10^6 M_\odot $  and distance $D=8$kpc from Earth, the EHT demonstrate that the Sgr A* images are consistent with the expected appearance of a Kerr black hole \citep{EventHorizonTelescope:2022exc,EventHorizonTelescope:2022xqj}. Furthermore, when  compared with the EHT results for  M87*, it exhibits consistency with the predictions of GR stretching across three orders of magnitude in central mass \citep{EventHorizonTelescope:2022xnr}.

Nevertheless, the current uncertainty in the measurement of spin and the relative deviation of quadrupole moments do not eliminate black holes arising in MTG \citep{Akiyama:2019cqa, Akiyama:2019fyp,Akiyama:2019eap,Cardoso:2019rvt,EventHorizonTelescope:2021dqv}. Sgr A* black hole shadow observational results have already been used to test the viabilities of various non-Kerr black holes \citep{Vagnozzi:2022moj,Khodadi:2022pqh,Kuang:2022ojj,Pantig:2022ely,He:2022opa,Uniyal:2022vdu,Ghosh:2022kit,Chen:2022lct,Khodadi:2022pqh,Uniyal:2022vdu}. Photons that originate in the deep gravitational fields of black holes form the images, and thereby carry imprints of the spacetime properties in the strong-field regime \citep{Jaroszynski:1997bw,Falcke:1999pj}. Therefore, the black hole shadows can be used to perform the strong-field gravitational tests and eventually to test the no-hair theorem \citep{Johannsen:2010ru,Cunha:2018acu,Baker:2014zba}. The horizon-scale images of supermassive black holes provide a theoretically new route for testing the theory of GR.  Interestingly, the rotating regular black hole shadow's radius decreases and distortion in shadow increases monotonically with the parameters $g$ apart from spin $a$, and also a violation of the no-hair theorem significantly alters the shadow's shape and size \citep{Bambi:2008jg,Johannsen:2010ru,Falcke:2013ola,Johannsen:2013rqa,Johannsen:2016uoh}.

Until recently, however, precision gravity tests with black holes were impossible.  Regardless, the recent observations of black holes have become a reality, e.g., the detection of gravitational waves by LIGO/Virgo \citep{LIGOScientific:2016aoc,
 LIGOScientific:2016lio,
 LIGOScientific:2019fpa}, the detection of relativistic effects in the orbits of stars around Sgr A* \citep{GRAVITY:2018ofz,
GRAVITY:2020gka} and the EHT observations of the supermassive black holes M87* and Sgr A* \citep{Akiyama:2019cqa}. 
The horizon-scale image of Sgr A* can help test the black hole physics and GR predictions more satisfactorily than the image of M87*. The detection of relativistic effects precisely realized the mass and distance of Sgr A* in stellar orbits \citep{Do:2019txf}. One can implement strong constraints on spacetime properties. Therefore, it would be interesting to use shadow observations of Sgr A* to distinguish rotating regular black holes/no-horizon spacetimes from the Kerr black hole/naked singularity in order to place constraints on the deviation parameters.  There is extensive evidence that Sgr A* contains a large amount of mass confined within a minimal volume. Whether it is an actual black hole with an event horizon remains unresolved. Sgr A* shadow observational results are consistent with the presence of an event horizon, proving that it ruled out all alternatives, e.g., naked singularities, and horizonless objects. Here, we discuss what EHT observations of Sgr A* can add to this question. 
If Sgr A* does not have an event horizon, it will probably have some surface \citep{EventHorizonTelescope:2022xnr,EventHorizonTelescope:2022xqj}. Alternatively, the object might be a boson star, naked singularity, or no-horizon spacetimes \citep{Cardoso:2019rvt}. Here, we test both rotating regular black holes and no-horizon spacetimes using Sgr A* observational data, and then the case for Sgr A* having an event horizon or being a no-horizon spacetime becomes significantly more robust.

The paper is organized as follows: In Section~\ref{sect2}, we discuss the null geodesic equations for most general stationary, axially symmetric regular black hole spacetimes as well as the mathematical technique for investigating the black hole shadow. Section~\ref{sect3} presents the shadows of four well-motivated rotating regular black holes, including those of their corresponding no-horizon spacetimes. Prior constraints on $(a,\, g)$ are also discussed. In Section~\ref{sect4} we numerically compute the shadow observables in order to compare the simulated shadows with the observed Sgr A* shadow. We derive constraints on the rotating regular black hole parameters with the aid of the Sgr A* black hole shadow. Finally, in Section~\ref{sect5} we summarize the obtained results and discuss the possibility of the candidature of rotating regular black holes for the astrophysical black holes.

\section{Analytic expressions for black hole shadows: General Rotating spacetimes }\label{sect2}
A black hole, placed in a bright background, casts a shadow that appears as a dark region in the observer's plane, edged by a characteristic sharp and bright ring, known as a photon ring, whose structure solely depends upon the spacetime geometry \citep{CT}. The black hole shadow results from the strong-field gravity near the event horizon, and it can in principle be used to determine the properties of the black hole spacetime, such as its spin \citep{Kumar:2018ple}. Hence, we investigate the photon geodesics around the rotating regular black hole spacetime, whose line element in Boyer$-$Lindquist coordinates ($t, r, \theta, \phi$) reads \citep{Bambi:2013ufa,Kumar:2018ple,Kumar:2020ltt}
\begin{eqnarray}\label{rotmetric}
ds^2 & = & - \left[ 1- \frac{2m(r)r}{\Sigma} \right] dt^2  - \frac{4am(r)r}{\Sigma  } \sin^2 \theta\, dt \, d\phi +
\frac{\Sigma}{\Delta}dr^2  \nonumber
\\ && + \Sigma\, d \theta^2+ \left[r^2+ a^2 +
\frac{2m(r) r a^2 }{\Sigma} \sin^2 \theta
\right] \sin^2 \theta\, d\phi^2,
\end{eqnarray}
where
$
\Sigma = r^2 + a^2 \cos^2\theta,\;\;\;\;\;  \Delta = r^2 + a^2 - 2m(r)r.$, $m(r)$ is the mass function with  $\lim_{r\to\infty}m(r)=M$ and $a$ is the spin parameter defined as $a=J/M$; $J$ and $M$ are, respectively, the angular momentum and ADM mass of rotating black hole. The metric (\ref{rotmetric}) goes over to Kerr (\citeyear{Kerr:1963ud}) and Kerr$-$Newman \citep{Newman:1965my} spacetimes, respectively, when $m(r)=M$ and $m(r)=M-Q^2/2r$. 
In the metric (\ref{rotmetric}), one can suitably replace $m(r)$ to get a wide range of rotating regular black hole spacetime  \citep{Ghosh:2014pba,Abdujabbarov:2016hnw,Amir:2016cen,Kumar:2018ple,Kumar:2020ltt,Islam:2021ful,Ghosh:2020spb}. 

The Hamilton$-$Jacobi equation determines the photon motion in the spacetime (\ref{rotmetric}) \citep{Carter:1968rr} 
\begin{eqnarray}
\label{HmaJam}
\frac{\partial S}{\partial \tau} = -\frac{1}{2}g^{\alpha\beta}\frac{\partial S}{\partial x^\alpha}\frac{\partial S}{\partial x^\beta} ,
\end{eqnarray}
The corresponding geodesics equations can be obtained from Eq.~(\ref{HmaJam}).  Here, $\tau$ is the affine parameter along the geodesics, and $S$ is the Jacobi action, given by
\begin{eqnarray}
S=-{\cal E} t +{\cal L} \phi +S_r(r)+S_\theta(\theta) \label{action},
\end{eqnarray}
$S_r(r)$ and $S_{\theta}(\theta)$, respectively, are functions only of the $r$ and $\theta$ coordinates. The metric (\ref{rotmetric}) is time translational and rotational symmetry, which leads to two conserved quantities, namely energy $\mathcal{E}=-p_t$ and axial angular momentum $\mathcal{L}=p_{\phi}$, where $p_{\mu}$ is the photon's four-momentum. 
The Petrov-type $D$ character of metric~(\ref{rotmetric}) guarantees the separable constant $\mathcal{K}$ \citep{Carter:1968rr}, and then we get a set of null geodesics equations in the first-order differential form \citep{Carter:1968rr,Chandrasekhar:1985kt}:
\begin{eqnarray}
\Sigma \frac{dt}{d\tau}&=&\frac{r^2+a^2}{\Delta}\left({\cal E}(r^2+a^2)-a{\cal L}\right)  -a(a{\cal E}\sin^2\theta-{\mathcal {L}})\ ,\label{tuch}\\
\Sigma \frac{dr}{d\tau}&=&\pm\sqrt{\mathcal{V}_r(r)}\ ,\label{r}\\
\Sigma \frac{d\theta}{d\tau}&=&\pm\sqrt{\mathcal{V}_{\theta}(\theta)}\ ,\label{th}\\
\Sigma \frac{d\phi}{d\tau}&=&\frac{a}{\Delta}\left({\cal E}(r^2+a^2)-a{\cal L}\right)-\left(a{\cal E}-\frac{{\cal L}}{\sin^2\theta}\right)\ ,\label{phiuch}
\end{eqnarray}
where 
$\mathcal{V}_r(r)$ and $\mathcal{V}_{\theta}(\theta)$, respectively, are related to the following effective potentials for radial and polar motion:   
\begin{eqnarray}\label{06}
\mathcal{V}_r(r)&=&\left[(r^2+a^2){\cal E}-a{\cal L}\right]^2-\Delta[{\cal K}+(a{\cal E}-{\cal L})^2]\label{rpot},\quad \\ 
\mathcal{V}_{\theta}(\theta)&=&{\cal K}-\left[\frac{{\cal L}^2}{\sin^2\theta}-a^2 {\cal E}^2\right]\cos^2\theta.\label{theta0}
\end{eqnarray}
The $\mathcal{K}$ is related to the Carter constant $\mathcal{Q}=\mathcal{K}+(a\mathcal{E}-\mathcal{L})^2$ (\citeyear{Carter:1968rr}), which is essentially a manifestation of the isometry of metric (\ref{rotmetric}) along the second-order Killing tensor field. To proceed further, we introduce two dimensionless impact parameters \citep{Chandrasekhar:1985kt} 
\begin{equation}
\xi\equiv \mathcal{L}/\mathcal{E},\qquad \eta\equiv \mathcal{K}/\mathcal{E}^2,
\end{equation}
to reduce the degrees of freedom of the photon geodesics from three to two. 
Depending on the critical parameters' values, the photon may either get captured, scatter to infinity, or form bound orbits, which depend on the radial effective potential $\mathcal{V}_r(r)$ and the constants of motion. It turns out that the unstable photon orbits outline the black hole shadow, which can be determined by 
\begin{equation}
\left.\mathcal{V}_r\right|_{(r=r_p)}=\left.\frac{\partial \mathcal{V}_r}{\partial r}\right|_{(r=r_p)}=0,\,\, \text{and}\,\, \left.\frac{\partial^2 \mathcal{V}_r}{\partial r^2}\right|_{(r=r_p)}> 0,\label{vr} 
\end{equation}
where $r_p$ is the radius of an unstable photon orbit.  Solving Eq.~(\ref{vr}) for Eq.~(\ref{rpot}) results in the critical impact parameters \citep{Abdujabbarov:2016hnw,Kumar:2018ple}
\begin{align}
\xi_c=&\frac{[a^2 - 3 r_p^2] m(r_p) + r_p [a^2 + r_p^2] [1 + m'(r_p)]}{a [m(r_p) + r_p [-1 + m'(r_p)]]},\nonumber\\
\eta_c=&-\frac{r_p^3}{a^2 [m(r_p) + r_p [-1 + m'(r_p)]]^2}\Big[r_p^3 + 9 r_p m(r_p)^2 \nonumber\\
&+ r_pm'(r_p)[4 a^2 + 2r_p^2+r_p^2 m'(r_p)]  \nonumber\\
&-  2 m(r_p) [2 a^2 + 3 r_p^2 + 3 r_p^2 m'(r_p)]\Big],\label{impactparameter}
\end{align}
where $'$ stands for the derivative with respect to the radial coordinate. Equation~(\ref{impactparameter}) reduces to that of the Kerr black hole when $m(r)=M$. In particular, photons with $\mathcal{\eta}_c=0$ form planar circular orbits confined only to the equatorial plane, whereas $\mathcal{\eta}_c>0$ leads to three-dimensional spherical orbits \citep{Chandrasekhar:1985kt}. In rotating spacetimes, photons can either have prograde motion or retrograde motion, whose respective radii at the equatorial plane, $r_p^{-}$ and $r_p^{+}$, can be identified as the real positive roots of $Y=0$ for $r_p\geq r_+$, and all other spherical photon orbits have radii of $r_p^-< r_p<r_p^+$. Further, the maximum latitude of spherical orbits depends on the angular momentum of photons, i.e., the smaller the angular momentum, the larger the orbit latitude. Photons must have zero angular momentum to reach the polar plane of the black hole, whose orbit radius $r_p^0$, such that $r_p^-\leq r_p^0\leq r_p^+$, can be determined by the zeros of $\xi_c=0$.  

The prograde and retrograde circular photon orbits radii, respectively, in the equatorial plane read \citep{Teo:2003} 
\begin{eqnarray}\label{photonRKerr}
r_p^-&=&2M\left[1+ \cos\left(\frac{2}{3}\cos^{-1}\left[-\frac{|a|}{M}\right]\right) \right],\nonumber\\
r_p^+&=&2M\left[1+ \cos\left(\frac{2}{3}\cos^{-1}\left[\frac{|a|}{M}\right]\right) \right],
\end{eqnarray}  
and they are, respectively, in the ranges $M\leq r_p^-\leq 3M$ and  $3M\leq r_p^+\leq 4M$. Indeed, for the Schwarzschild black hole ($a=0$), these two circular orbits coincide into a single orbit, and the photon region in Eq.~(\ref{photonRKerr}) degenerates into a photon sphere of constant radius $r_p=3M$. The $r_p^-\leq r_p^+$ can be attributed to the Lense-Thirring effect \citep{Johannsen:2010ru}, whereas the prograde orbits and event horizon radii coincide with $r_p^-=r_E=M$  for the extremal Kerr black hole ($a=M$).

Next, we derive the relation between two constants, $\xi_c$ and $\eta_c$, and the observer's image plane coordinates, $X$ and $Y$. For an observer at the position ($r_o,\theta_o$) and using the tetrad components of the four-momentum $p^{(\mu)}$ and geodesic Eqs.~(\ref{tuch}), (\ref{th}), and (\ref{phiuch}), we obtain:
\begin{eqnarray}
&&X= -r_o\frac{p^{(\phi)}}{p^{(t)}} = -\left. r_o\frac{\xi_c}{\sqrt{g_{\phi\phi}}(\zeta-\gamma\xi_c)}\right|_{(r_o,\theta_o)},\nonumber\\
&&Y = r_o\frac{p^{(\theta)}}{p^{(t)}} =\pm\left. r_o\frac{\sqrt{\mathcal{V}_{\theta}(\theta)}
}{\sqrt{g_{\theta\theta}}(\zeta-\gamma\xi_c)}\right|_{(r_o,\theta_o)},~\label{Celestial}
\end{eqnarray} 
where 
\begin{eqnarray}
\zeta=\sqrt{\frac{g_{\phi\phi}}{g_{t\phi}^2-g_{tt}g_{\phi\phi}}},\qquad \gamma=-\frac{g_{t\phi}}{g_{\phi\phi}}\zeta.
\end{eqnarray}
where the coordinates $X$ and $Y$ in Eq.~(\ref{Celestial}), respectively, denote the apparent displacement along the perpendicular and parallel axes to the projected axis of the black hole symmetry. For an observer in an asymptotically flat region ($r_o\to\infty$), the celestial coordinate Eq.~(\ref{Celestial}), yields  \citep{bardeen1973}
\begin{equation}
X=-\xi_c\csc\theta_o,\qquad Y=\pm\sqrt{\eta_c+a^2\cos^2\theta_o-\xi_c^2\cot^2\theta_o}.\label{pt}
\end{equation} 
The parametric plots of Eqs.~(\ref{Celestial}) or (\ref{pt}) in the ($X, Y$) plane can cast a variety of black hole shadows for different choices of black hole mass function $m(r)$ \citep{Abdujabbarov:2016hnw,Amir:2016cen,Kumar:2019pjp,Kumar:2018ple,Kumar:2020ltt}, and when  $m(r)=M$, we get the shadows of the Kerr black holes \citep{bardeen1973,Kumar:2018ple}.

\begin{figure*}
    \begin{tabular}{c c}
    \includegraphics[scale=0.85]{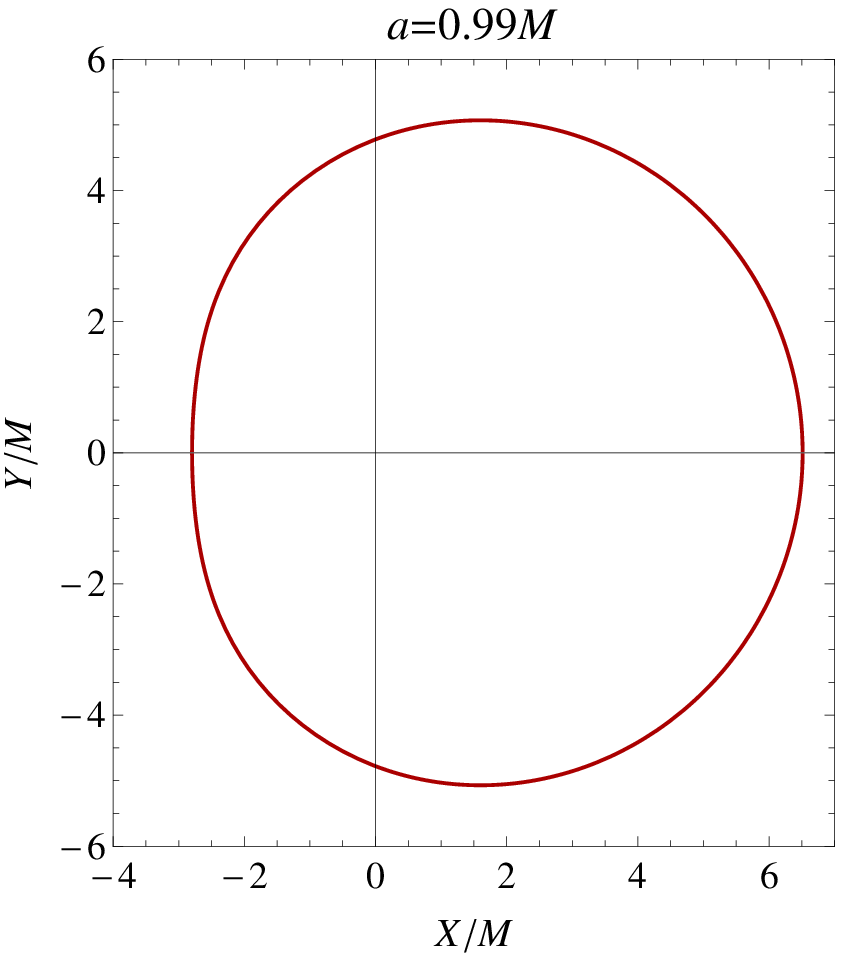} & 
    \includegraphics[scale=0.85]{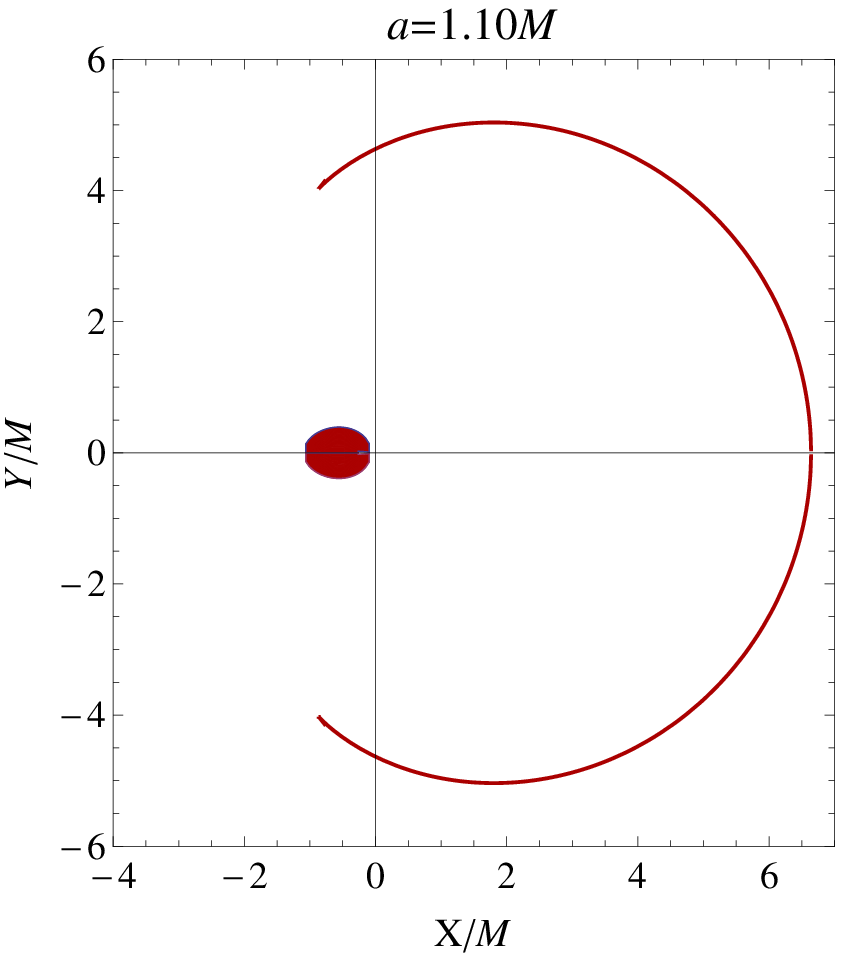}
    \end{tabular}
    \caption{Shadow silhouette for a Kerr black hole (left) and the Kerr naked singularity (right) for an inclination angle $\theta_0=50^{o}$. Kerr naked singularity ($a>M$) always cast open arc shadow.}
    \label{fig:KerrShadow}
\end{figure*}

The apparent shape of a photon ring is perfectly circular for all values of $a$ and inclination angle  $\theta_o=0, \pi$. Indeed, the Kerr black holes ($a\leq M$) always admit both prograde and retrograde unstable photon orbits and form a closed photon ring or shadow silhouette \citep{Kumar:2020yem,Kumar:2020ltt}. However, for a Kerr naked singularity $a>M$, prograde photons on the equatorial plane and a few neighboring spherical photon orbits spiral in and end up in the central singularity. The retrograde photon orbits cause an open arc ring with a tiny segment missing near the left endpoint \citep{Charbulak:2018wzb,Hioki:2009na}. Figure~\ref{fig:KerrShadow} shows that the photon ring structure for the Kerr naked singularity is markedly different from that of a Kerr black hole. For a Kerr black hole, the shape of this photon ring is nearly circular unless the black hole is rapidly spinning. 
According to the cosmic censorship conjecture (CCC), the event horizon hides the black hole singularity, forbidding naked singularities \citep{Penrose:1969pc}. The proof of CCC's validity is still an open problem in relativity, and CCC remains unproven to date.  Indeed, the closed photon ring in the Kerr spacetime is in accordance with the CCC; the photon ring in Kerr spacetime is closed as long as the horizon is present. The naked singularity of spherically symmetric Janis-Newman-Winicour \citep{Shaikh:2019hbm} and Joshi-Malafarina-Narayan \citep{Shaikh:2018lcc} spacetimes, for some values of parameters, possess photon spheres and cast shadows similar to the Schwarzschild black hole shadow. However, in the absence of photon spheres, the images of naked singularities significantly differ from those of black holes \citep{Shaikh:2019hbm}. 
Therefore, it would be interesting to use shadow observations of Sgr A* to distinguish rotating regular black holes/no-horizon spacetimes from the Kerr black hole/naked singularity in order to place constraints on the deviation parameters. In the next section, we consider models involving four well-motivated rotating regular spacetimes, as well as other exotic possibilities, e.g., corresponding rotating regular no-horizon spacetimes. Hereafter, we consider the inclination angle $\theta_0=50^o$, which is relevant for the Sgr A* black hole \citep{EventHorizonTelescope:2022exc}.

\begin{table}
\begin{tabular}{|l||l|l|}
\hline
Rotating Regular Spacetimes & $g_E/M$  & $g_p/M$  \\ \hline\hline
Bardeen                     & 0.763332 & 0.82472 \\ \hline
Hayward                     & 1.05231  & 1.17566  \\ \hline
Ghosh-Culetu                & 1.202    & 1.2291  \\ \hline
Simpson-Visser              & 1.9950   & 2.9083   \\ \hline
\end{tabular}
\caption{The values of $g_E$ and $g_p$ for the rotating regular spacetimes ($a=0.1M$) and $\theta_0=50^{o}$. While the horizon disappears for $g>g_E$, the closed photon ring exists even for no-horizon spacetime with $g_E< g\leq g_p$.}
\label{Table1}
\end{table}

\section{Regular black holes and no-horizon spacetimes}\label{sect3}
\begin{figure*}
\begin{tabular}{c c }
	\includegraphics[scale=0.75]{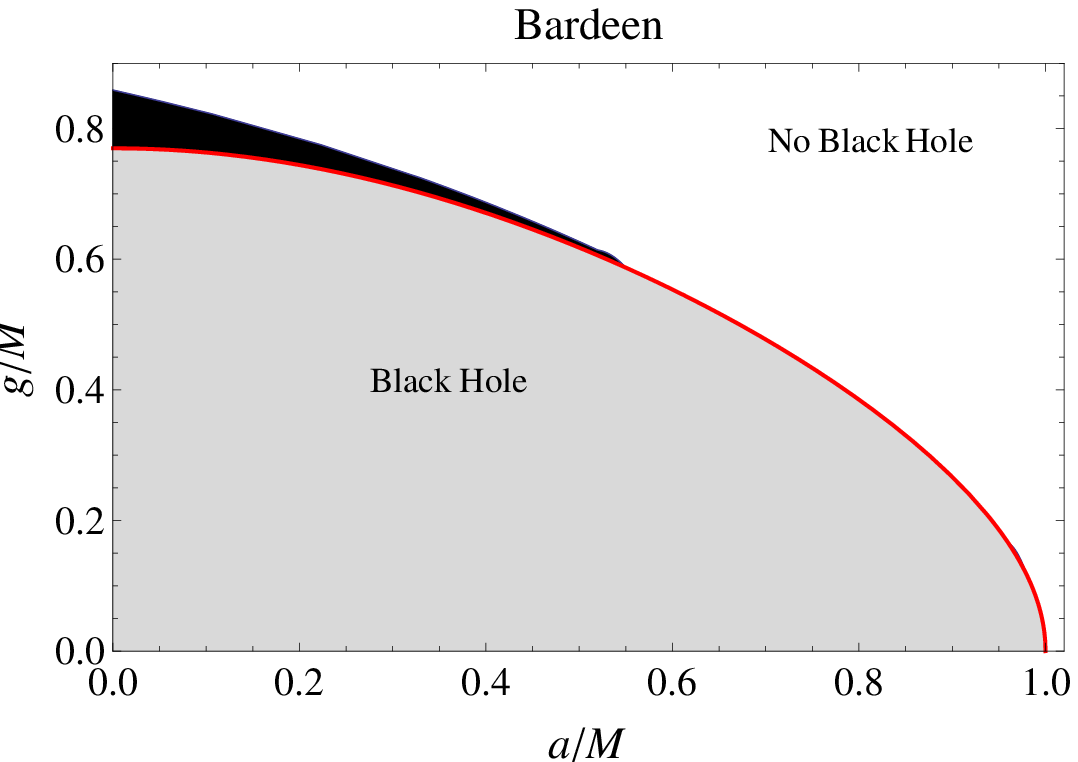}&
	\includegraphics[scale=0.75]{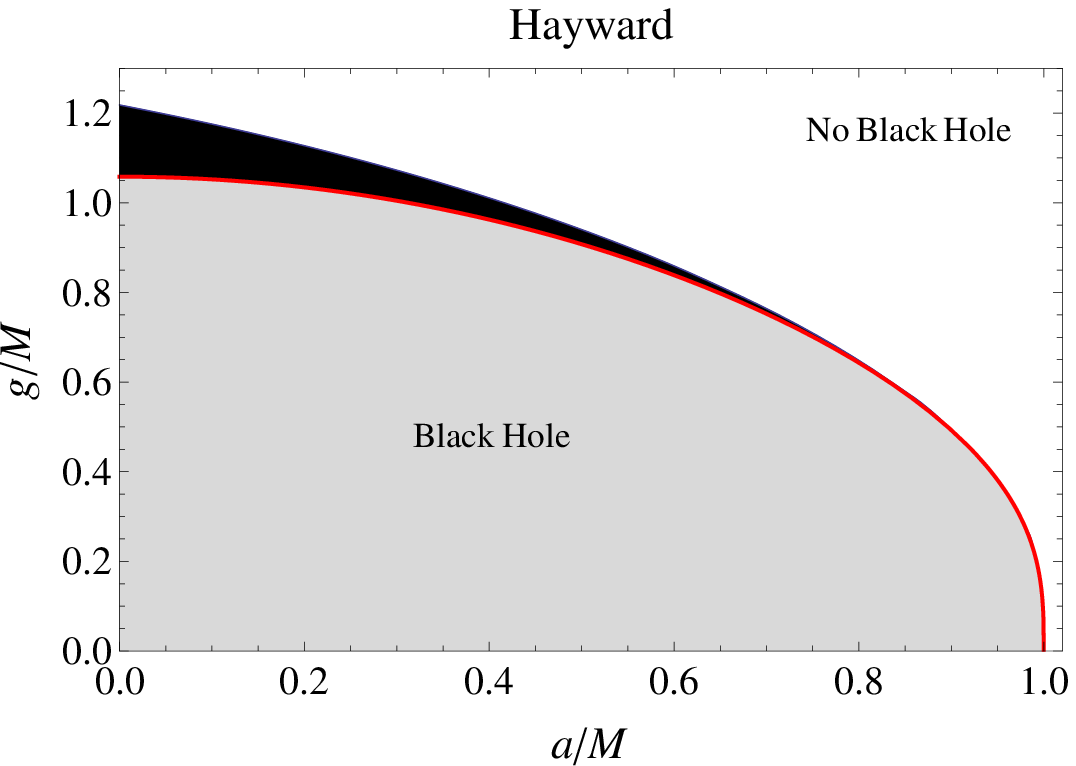}\\
	\includegraphics[scale=0.75]{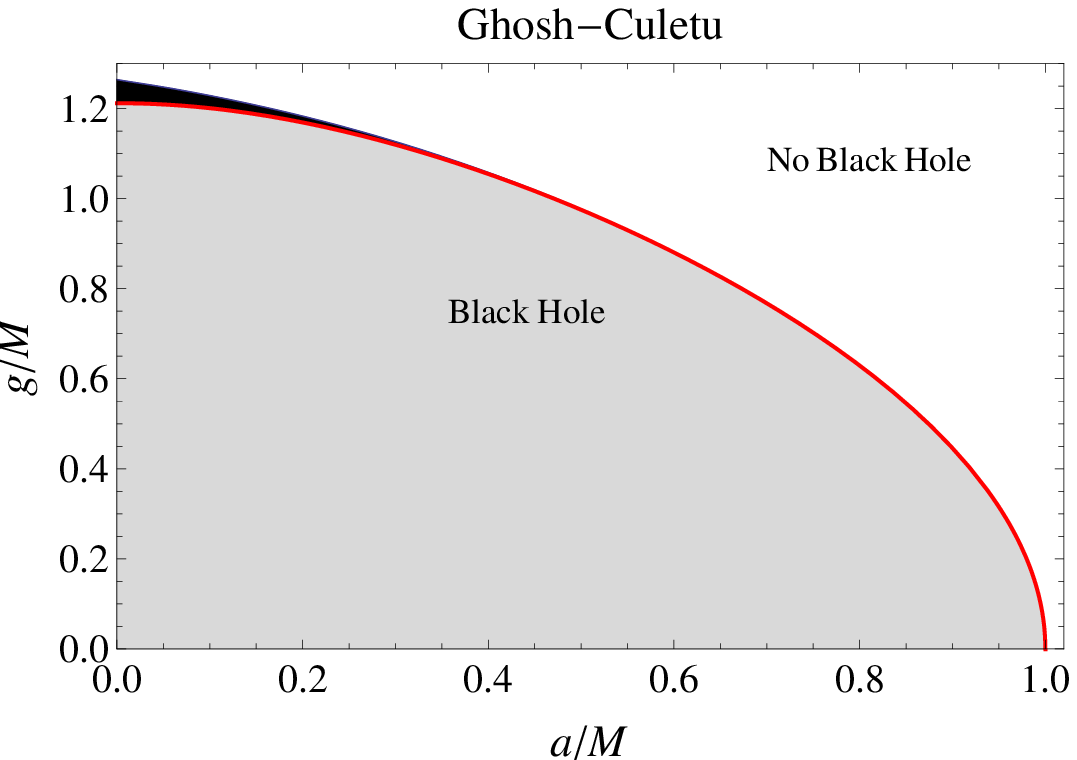}&
	\includegraphics[scale=0.75]{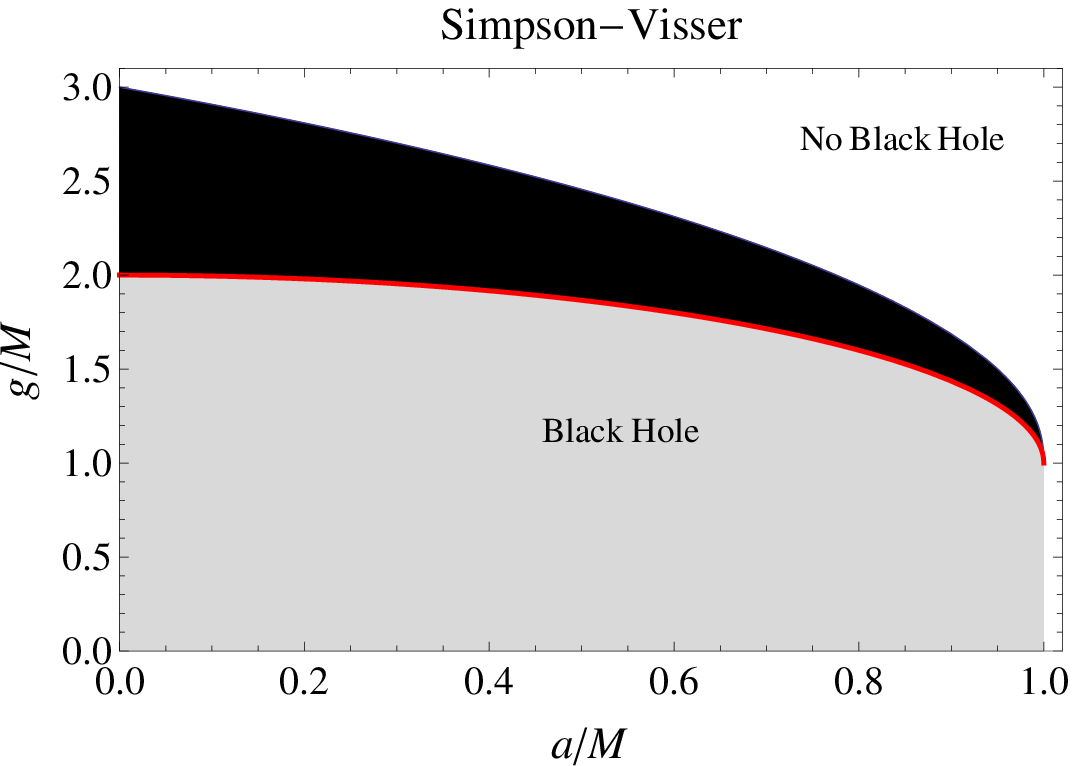}
\end{tabular}	
	\caption{Parameter space ($a, g$) corresponding to a black hole and no-horizon spacetime. Parameters in the grey regions define a black hole such that regions outside the red line represent no-horizon spacetime. No-horizon spacetime with parameters only in the black shaded region leads to the closed photon ring structure} \label{fig:BHNoBH}
\end{figure*}
Here, we examine four well-known rotating regular black holes, namely Bardeen \citep{Bambi:2013ufa,Kumar:2020yem,Kumar:2020ltt}, Hayward \citep{Bambi:2013ufa,Kumar:2020yem,Kumar:2020ltt}, Ghosh-Culetu \citep{Ghosh:2014pba} and Simpson-Visser black holes \citep{Mazza:2021rgq,Islam:2021ful}. The regularity of these black holes spacetimes are investigated in terms of the curvature scalars, which are finite everywhere including at the center, as shown in Appendix \ref{sec:appendix}. It is important to note that the Hawking-Penrose singularity theorem talks about the spacetime singularities in terms of the geodesics incompleteness rather than the singularity of curvature invariant \citep{Hawking:1971vc}. Thus, the ultimate proof of regularity is the geodesic completeness. Nevertheless, for $r\geq 0$, these regular black holes are geodesically complete \citep{Zhou:2022yio}.  
The shadows of these regular models have received significant attention and their shape and size are considerably different from those of the Kerr black hole shadows \citep{Abdujabbarov:2016hnw,Amir:2016cen,Kumar:2019pjp,Kumar:2018ple,Kumar:2020ltt}. 
The rotating regular spacetimes in question belong to a family of prototype non-Kerr black hole metrics with an additional deviation parameter of $g$ related to the nonlinear electrodynamics charge besides the $a$ and $M$ of the Kerr black hole, which is included as a special case of vanishing deviation parameter $g=0$. Interestingly, for a given $ a $, there is a critical value of $ g $, $g_E$ such that $\Delta=0$  has no zeros for $ g >  g_E$ and one double zero at $ r = r_E  $ for $ g =  g_E $, respectively corresponding to a no-horizon regular spacetime and an extremal black hole with degenerate horizons. We show that, unlike the Kerr naked singularity, no-horizon regular spacetimes can possess a closed photon ring when  $g_E< g \leq g_p$ (cf. Fig.~\ref{fig:BHNoBH} and Table \ref{Table1}) \citep{Kumar:2020ltt}. Interestingly, the value of $g_p$ depends on the inclination angle $\theta_0$.

The EHT collaboration released the first image of the supermassive black hole M87* in 2019, delivering an emission ring of diameter $\theta_d= 42\pm 3\,\mu$as \citep{Akiyama:2019cqa,Akiyama:2019eap,Akiyama:2019fyp}. In Ref.~\citep{Kumar:2020yem}, we have shown that the shadows of Bardeen black holes ($g\lesssim 0.26 M$), Hayward black holes ($g\lesssim 0.65 M$), Ghosh-Culetu black holes ($g\lesssim 0.25 M$) and Simpson-Visser (independent of $g$) are indistinguishable from Kerr black hole shadows within the current observational uncertainties. Thus, we can consider these black holes as powerful, viable candidates for astrophysical black holes. Furthermore, Bardeen black holes ( $g\leq 0.30182M$), Hayward black holes ($g\leq 0.73627M$), and Ghosh-Culetu black holes ($g\leq 0.30461M$), within the $1\sigma$ region for $\theta_d= 39\, \mu$as,  are consistent with the observed angular diameter of M87* \citep{Kumar:2020yem}. In addition, unlike the Kerr naked singularity, no-horizon regular spacetimes can possess a closed photon ring when  $g_E< g \leq g_p$, e.g., for $a=0.10M$, Bardeen  ($g_E=0.763332M<g\leq g_p= 0.816792M$), Hayward ($g_E=1.05297M < g\leq g_p = 1.164846M$) and Ghosh-Culetu ($g_E=1.2020M < g \leq g_p= 1.222461M$) no-horizon spacetimes have closed photon ring (cf. Fig. \ref{fig:BHNoBH} and Table \ref{Table1}) \citep{Kumar:2020ltt}.

\begin{figure*}
\begin{tabular}{c c }
	\includegraphics[scale=0.8]{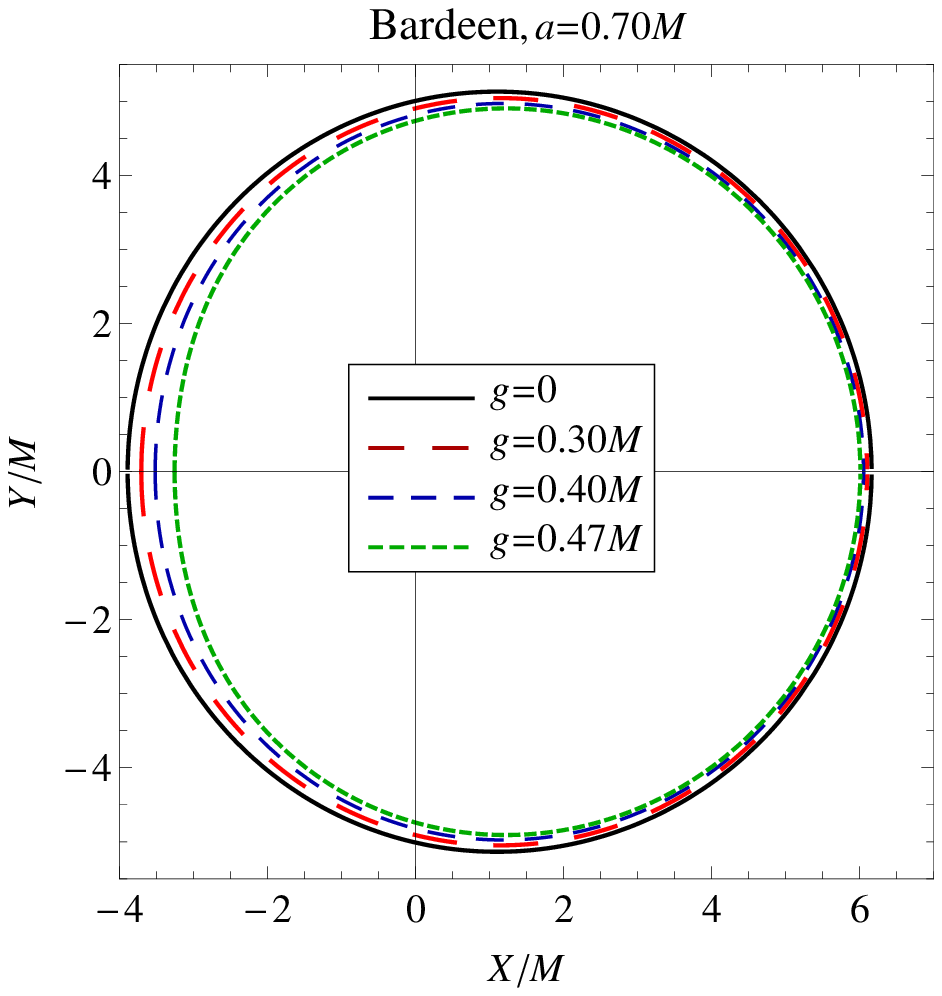}&
	\includegraphics[scale=0.8]{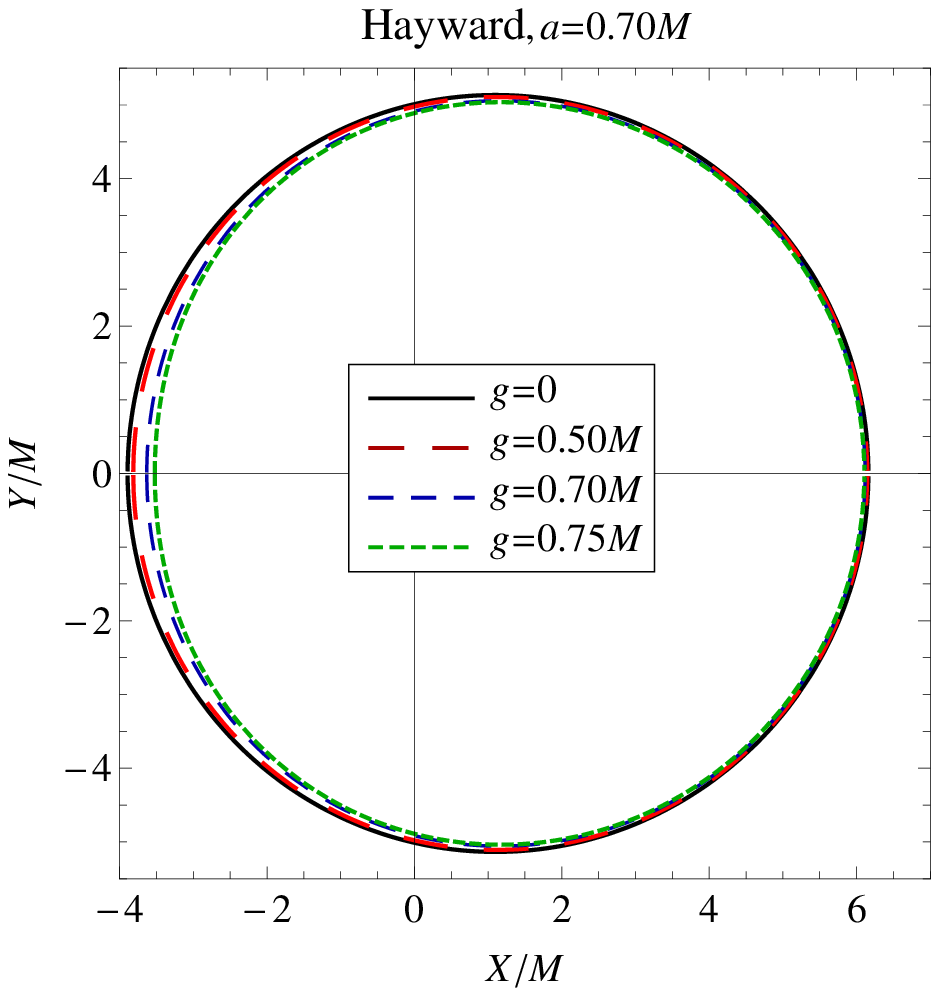}\\
	\includegraphics[scale=0.8]{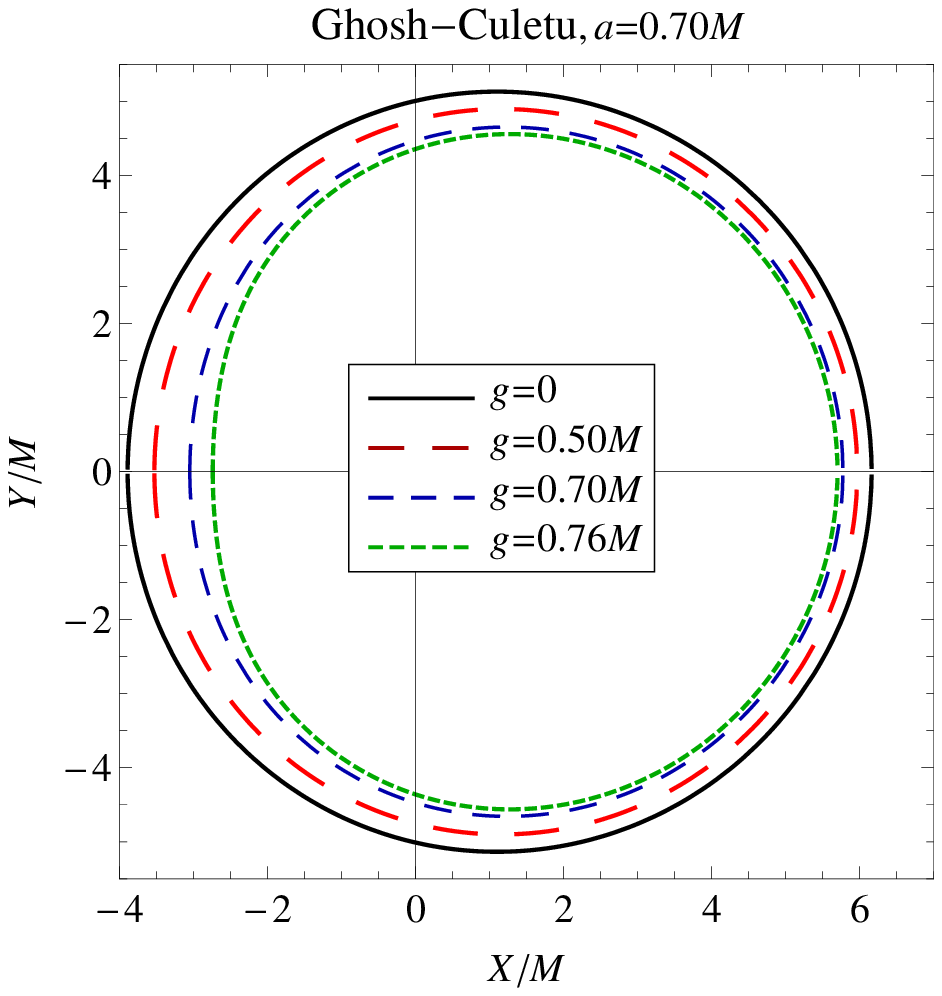}&
	\includegraphics[scale=0.8]{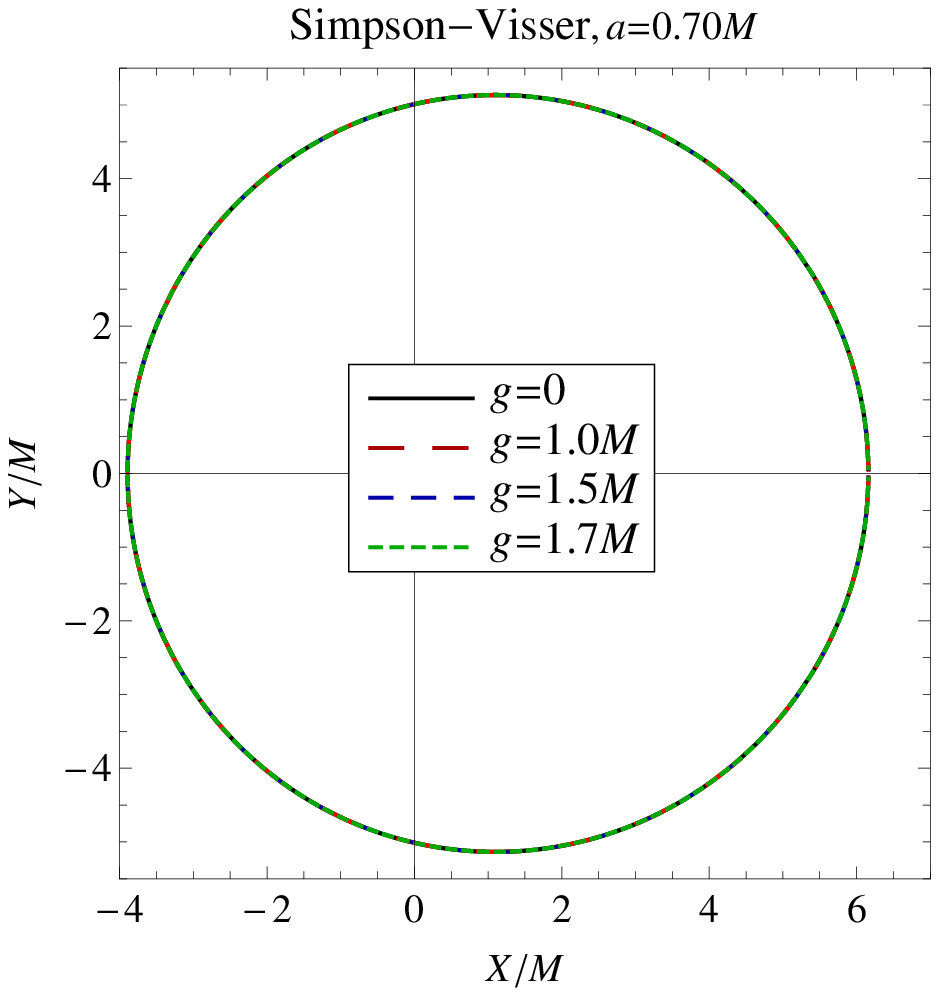}
\end{tabular}
	\caption{Regular black hole shadows silhouette with varying charge $g$. }\label{fig:BHShadow}
\end{figure*}
\begin{figure*}
\begin{tabular}{c c c c}
	\includegraphics[scale=0.8]{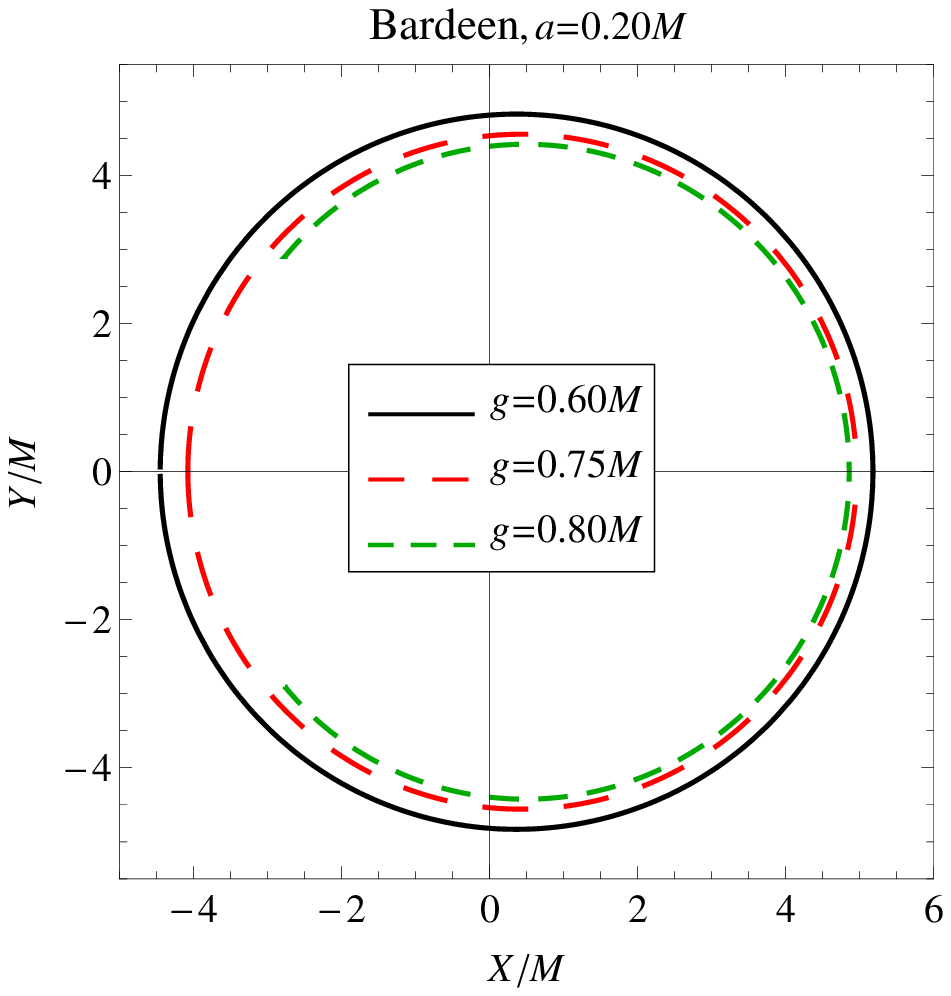}&
	\includegraphics[scale=0.8]{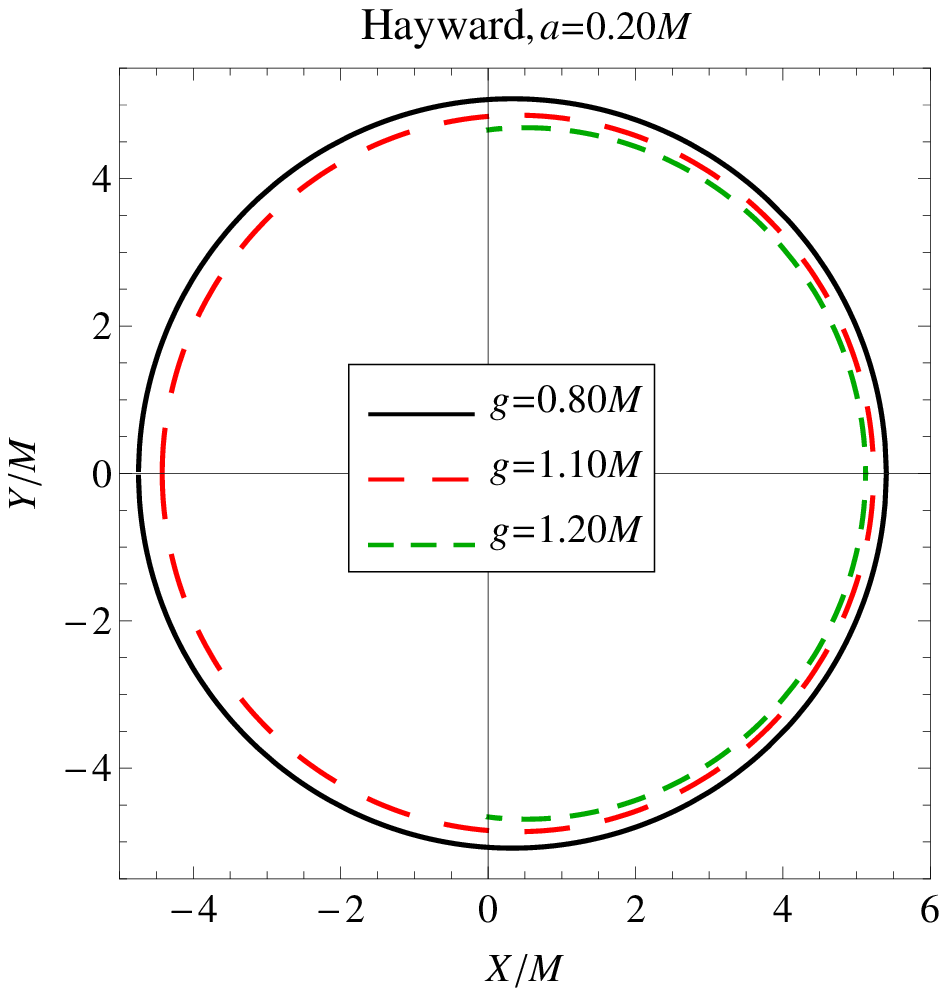}\\
	\includegraphics[scale=0.8]{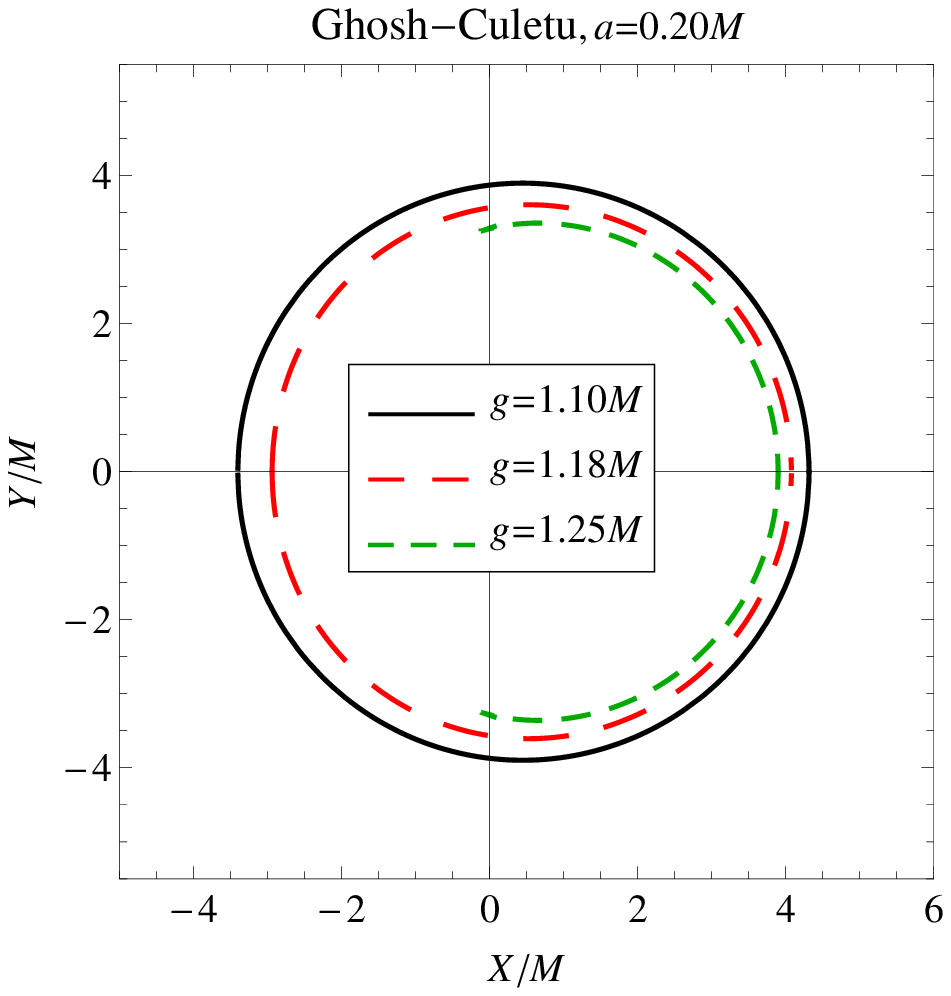}&
	\includegraphics[scale=0.8]{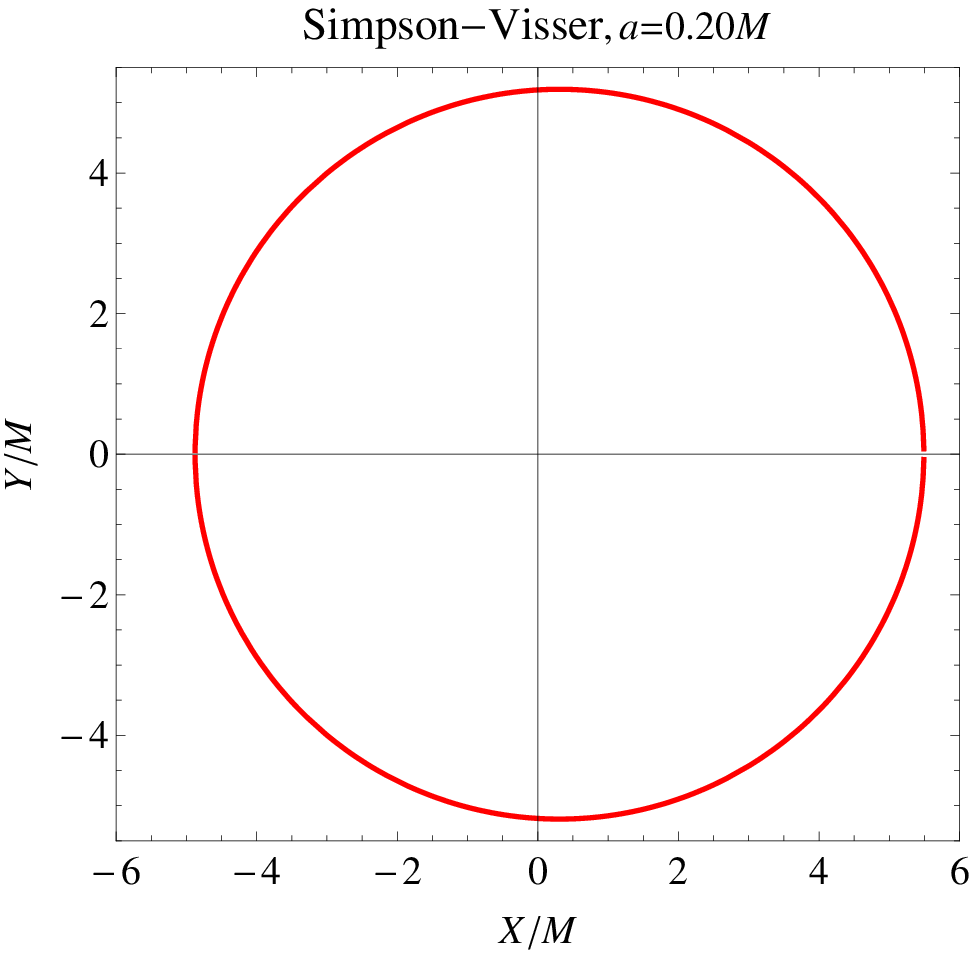}
\end{tabular}
	\caption{Figure showing a comparison of the silhouette of regular black holes and no-horizon spacetime shadows with varying charge $g$. The black curve represents the black hole case $g\leq g_E$, whereas the red and green curves are for the no-horizon spacetime $g\geq g_E$. Interestingly, no-horizon spacetime with $g_E\leq g\leq g_p$ casts a closed shadow boundary, as shown with the red  curve. }\label{fig:NoBHShadow}
\end{figure*}

\subsection{Bardeen Spacetime}\label{sec:bar}
\begin{figure*}
\begin{tabular}{c c}
	\includegraphics[scale=0.75]{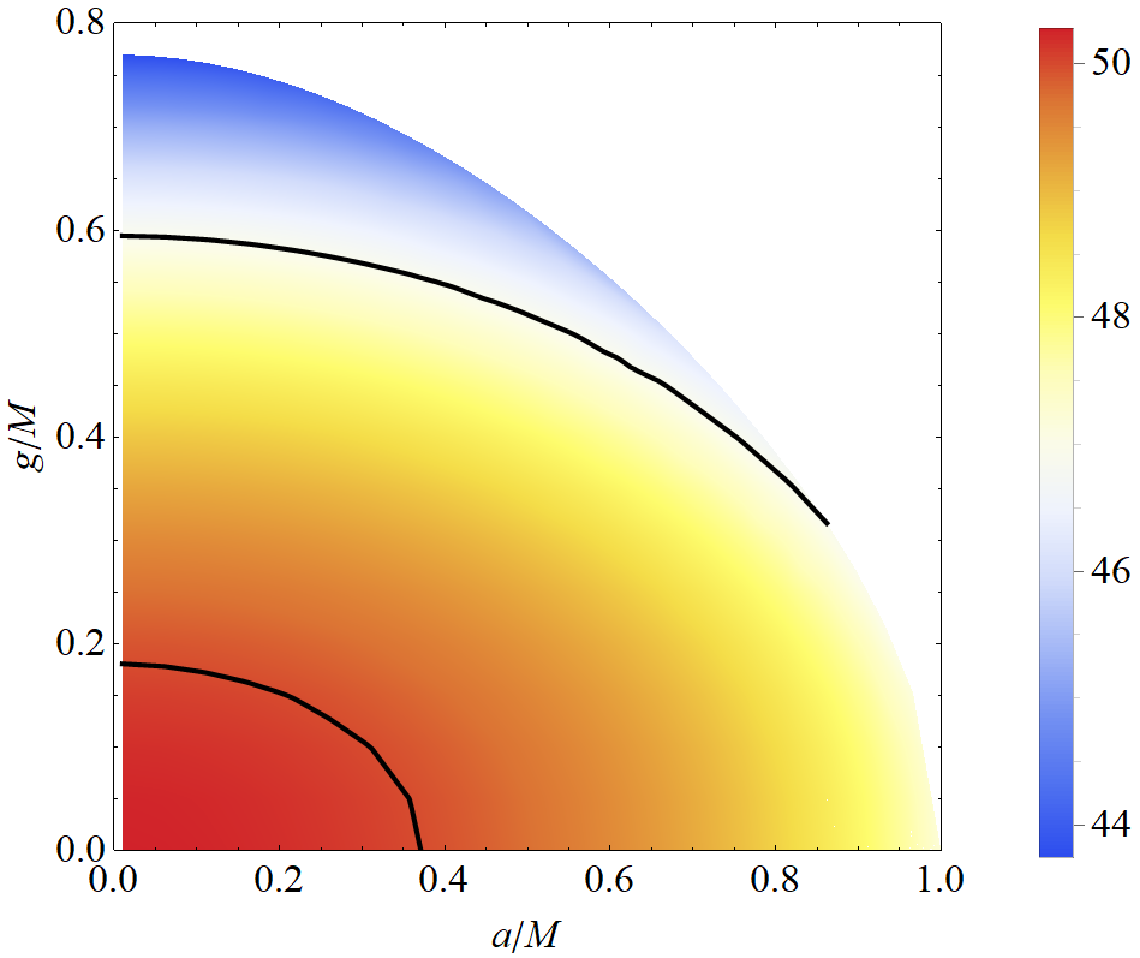}&
	\includegraphics[scale=0.75]{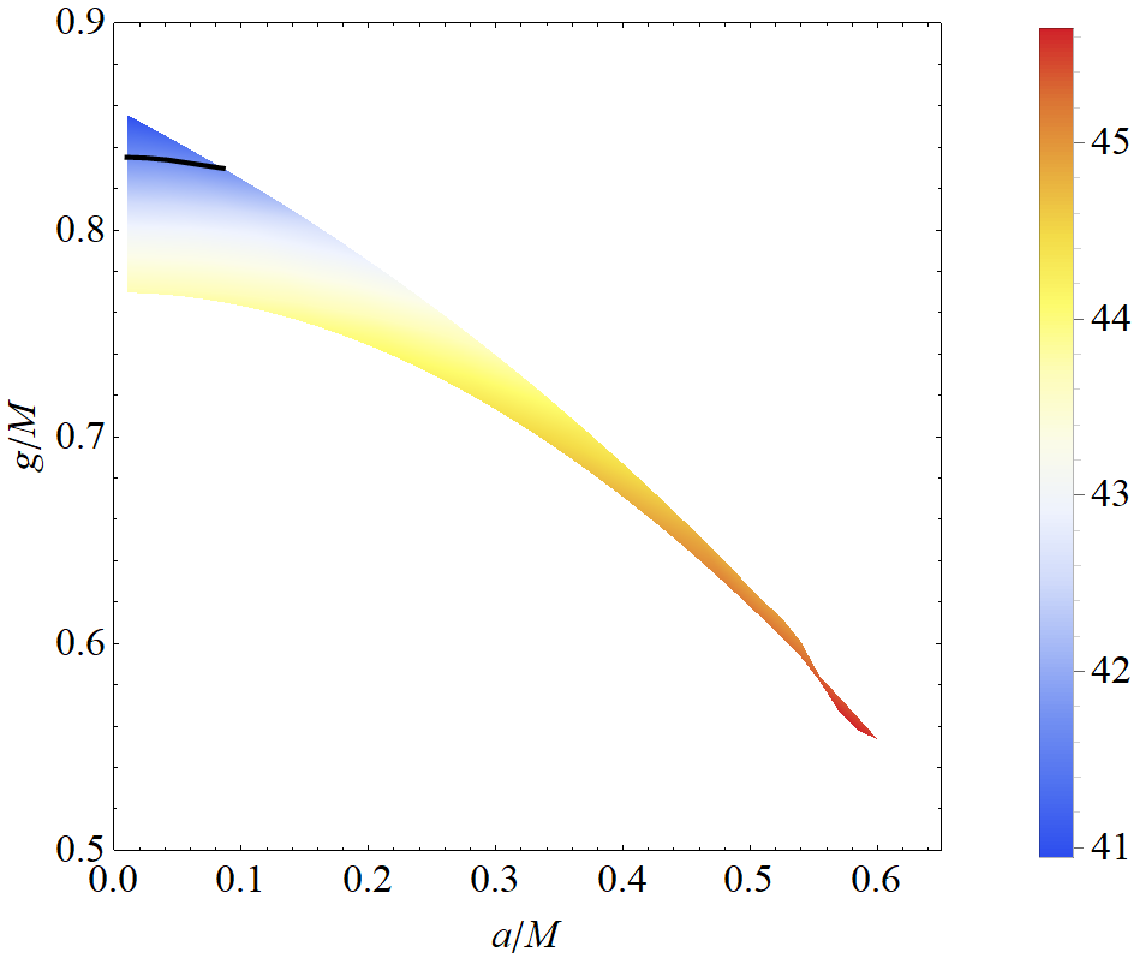}
\end{tabular}	
	\caption{(Left:) Bardeen black holes' shadow angular diameter $\theta_{sh}$, in units of $\mu$as, as a function of ($a, g$). Black lines are for $46.9\mu$as and $50\mu$as corresponding to the Sgr A* black hole shadow bounds. Bardeen black holes with parameter region between these two black lines cast shadows that are consistent with the Sgr A* shadow size. (Right:) Bardeen no-horizon spacetime shadow angular diameter $\theta_{sh}$, in units of $\mu$as, as a function of ($a, g$). The black line is for $41.7\mu$as corresponding to the Sgr A* black hole shadow $1\sigma$ bounds, such that the region under the black line satisfies the Sgr A* shadow $1\sigma$ bound.}\label{fig:BarAng}
\begin{tabular}{c c}
	\includegraphics[scale=0.75]{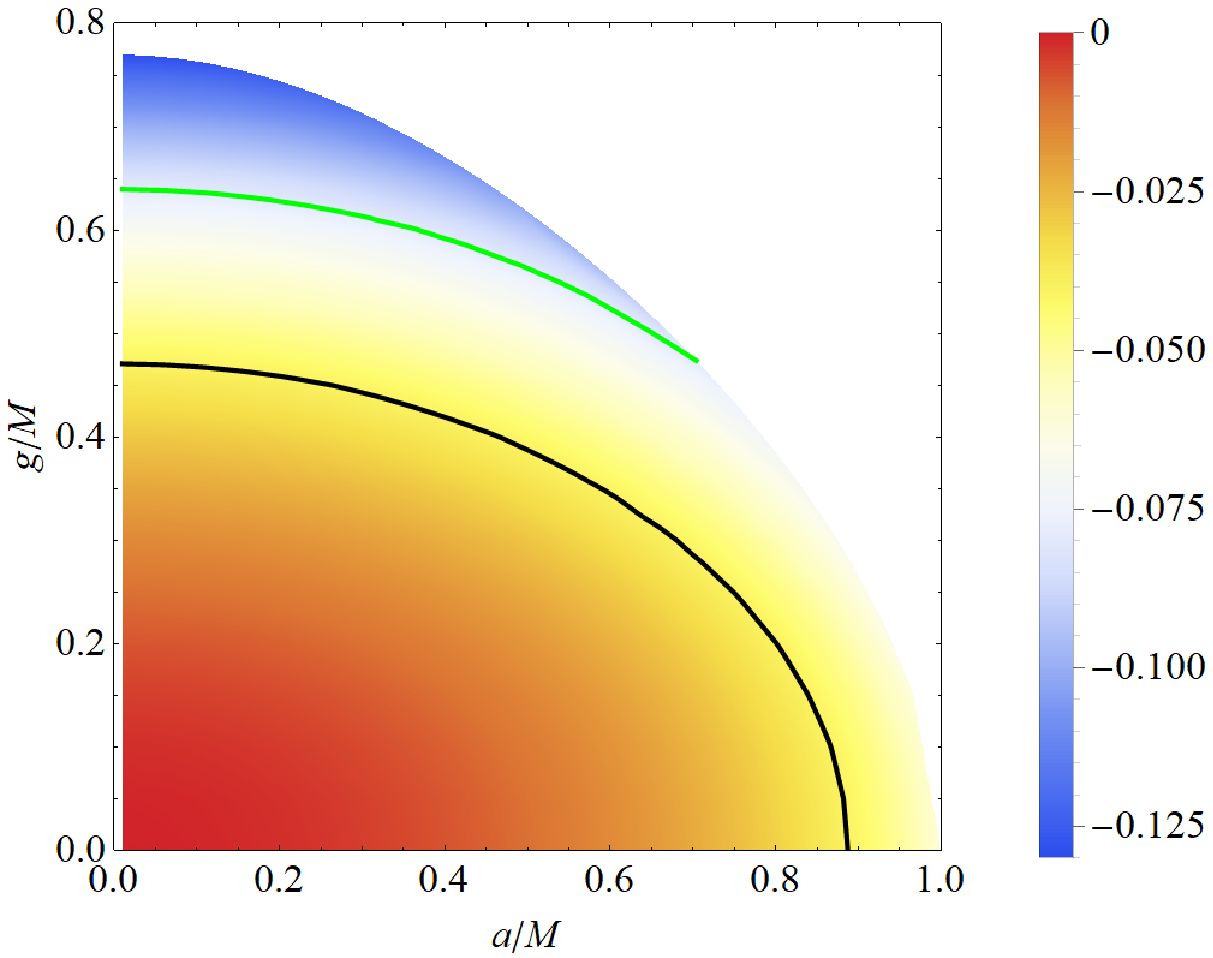}&
	\includegraphics[scale=0.75]{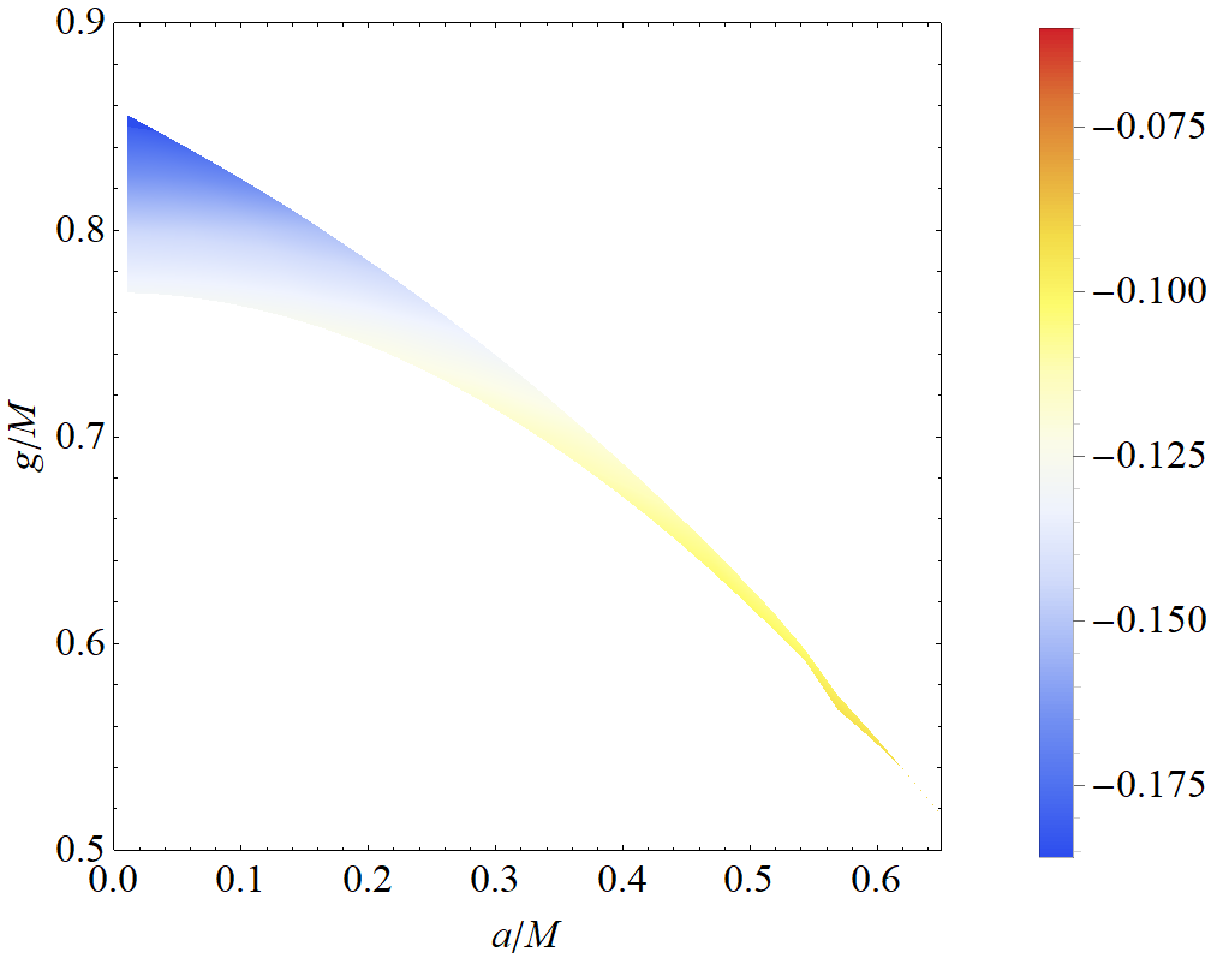}	
\end{tabular}	
	\caption{Bardeen black holes (left) and  no-horizons spacetime (right) shadows angular diameter deviation from that of a Schwarzschild black hole as a function of ($a, g$). The black and green lines are, respectively, for the  $\delta=-0.04$ (Keck)  and $\delta=-0.08$ (VLTI) corresponding to the Sgr A* black hole shadow bounds. }\label{fig:BarDelta}
\end{figure*}

An idea first put forth by Sakharov \citep{Sakharov:1966aja} suggests that matter could avoid singularities, i.e., with a de Sitter core having an equation of state $P=-\rho$. Based on this idea, Bardeen (\citeyear{bardeen1968non}) proposed the first regular black hole solution, which has horizons but no curvature  singularity and is asymptotically ($r\to\infty$) flat. The rotating regular Bardeen black hole \citep{Bambi:2013ufa,Kumar:2020ltt,Kumar:2018ple,Kumar:2020yem,Ghosh:2015pra} is described by metric (\ref{rotmetric}) with the mass function \citep{bardeen1968non} 
\begin{equation}
m(r)=M\left(\frac{r^2}{r^2 + g^2}\right)^{3/2}, \label{Bardeenmass}
\end{equation}
where the deviation parameter $g$ can be identified as the magnetic monopole charge \citep{AyonBeato:2000zs} and at the limit $g\to 0$, one gets a Kerr black hole. First, the rotating Bardeen black hole shadows ($g\neq0$) become smaller and more distorted with increasing $g$ when compared with the Kerr black hole shadows (cf. Fig. \ref{fig:BHShadow} ) \citep{Abdujabbarov:2016hnw,Kumar:2020yem,Kumar:2020ltt,Tsukamoto:2014tja}.
Utilizing X-ray data from the disk surrounding the black hole candidate in Cygnus X-1, the rotating Bardeen black hole metric has been put to the test, such that the $3\sigma$ bounds $a_K>0.95 M$ \citep{Gou:2011nq} and $a_K>0.983M$ \citep{Gou:2013dna} for the Kerr metric infer respective bounds on Bardeen black hole parameters, respectively, $a>0.78M$ and $g<0.41M$, and $a>0.89M$ and $g<0.28M$ \citep{Bambi:2014nta}.
Although for $a=0$, the black hole horizons disappear for $g> g_E= 0.7698M$, the prograde and retrograde photon orbits exist for $g\leq 0.85865M$, resulting in a closed photon ring structure. In Figs.~\ref{fig:BHShadow} and \ref{fig:NoBHShadow}, we have shown the black holes and corresponding no-horizon shadow variation with varying charge $g$. Black hole shadow size decreases with both $a$ and $g$. For the Sgr A* model, the shadow angular diameter falls in $(43.7792, 50.2831)\mu$as, whereas the no-horizon spacetime shadow diameter is $\theta_{sh}\geq 40.7098\mu$as (cf. Fig.~\ref{fig:BarAng}). 

\subsection{Hayward Spacetime}\label{sec:hay}
Another thoroughly studied regular black hole model was proposed by Hayward (\citeyear{Hayward:2005gi}). Besides the mass $M$ it has one additional parameter $\ell$, which determines the length associated with the region concentrating the central energy density, such that modifications in the spacetime metric appear when the curvature scalar becomes comparable with $\ell^{-2}$. The spherically symmetric Hayward black hole model is identified as an exact solution of the general relativity minimally coupled to NED with magnetic charge $g$, where $g$ is related to $\ell$ via $g^3=2M\ell^2$ \citep{Fan:2016hvf}. The rotating Hayward black hole is also described by the metric (\ref{rotmetric}) with the mass function \citep{Hayward:2005gi,Kumar:2018ple}
\begin{equation}
m(r)=\frac{Mr^3}{r^3+g^3}.\label{Haywardmass}
\end{equation}

\begin{figure*}[ht]
\begin{tabular}{c c}
	\includegraphics[scale=0.75]{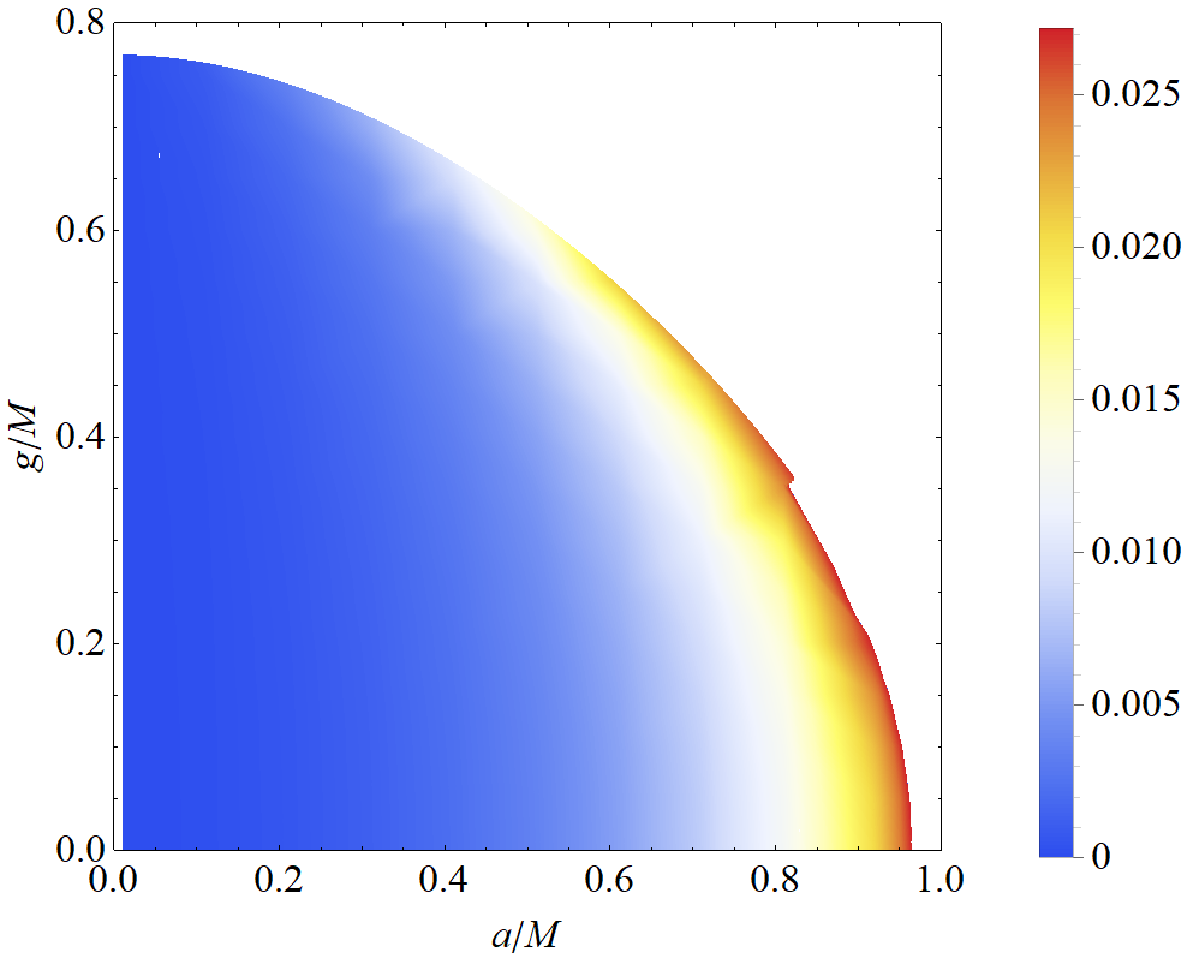}&
	\includegraphics[scale=0.75]{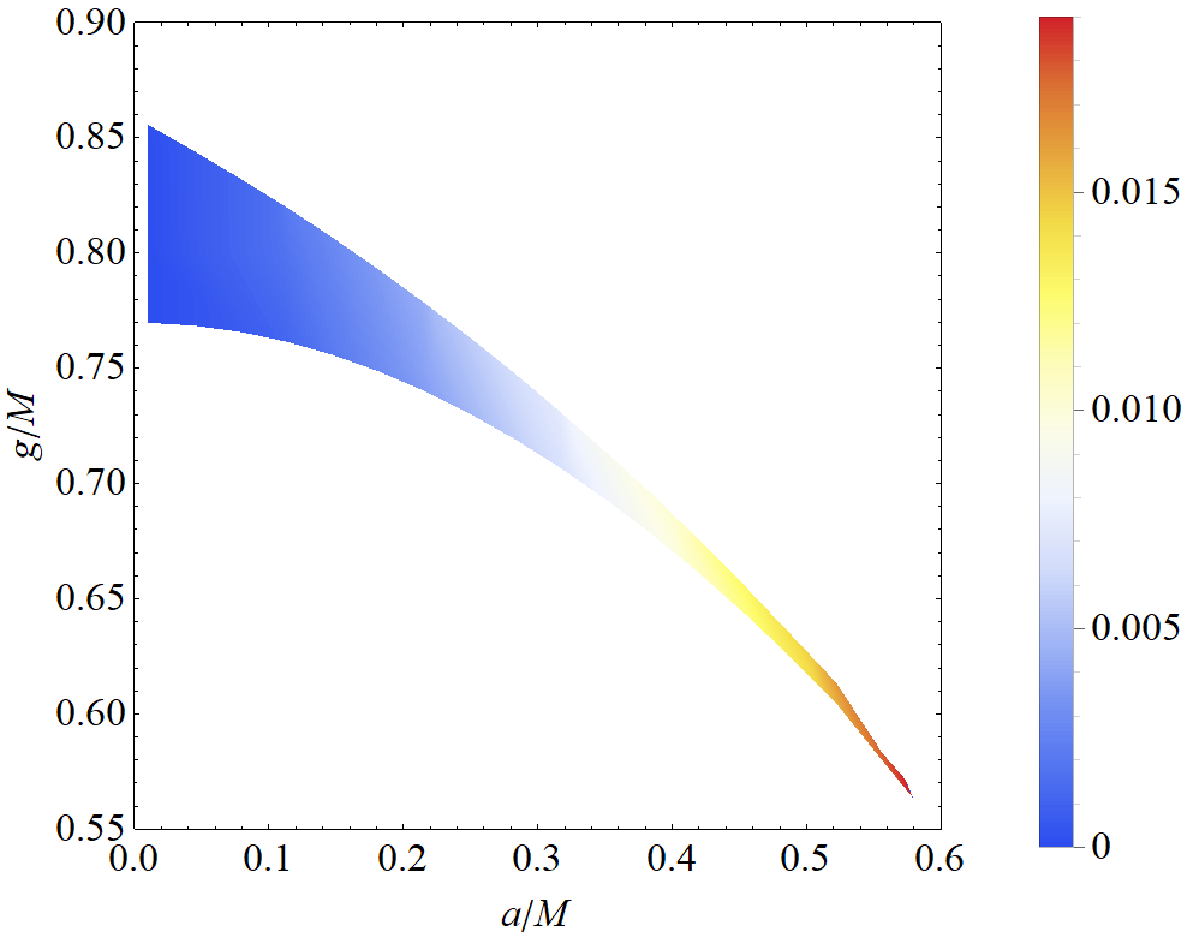}
\end{tabular}	
	\caption{Bardeen black holes (left) and no-horizons spacetime (right) shadows circularity deviation observable $\Delta C$ as a function of ($a, g$). }\label{fig:BarAsymm}
\end{figure*}
The photon region and shadows of rotating Hayward black holes have been extensively discussed \citep{Abdujabbarov:2016hnw, Kumar:2019pjp,Liu:2017ifc}.
It turns out that the bound $\Delta C \leq 0.1$  for the measured asymmetry of M87* shadow constrains $g \leq 1.0582M$ for the Hayward black holes \citep{Kumar:2020yem}. The angular diameter $\theta_{sh} = 39.6192\; \mu$as, which falls within the $1 \sigma$ confidence region with the observed angular diameter of the EHT observation of M87* black hole, strongly constrains the parameter $g$, i.e., $g \leq 0.73627M$ for the Hayward black holes \citep{Kumar:2020yem}. 
Although for $a=0$, the black hole horizons disappear for $g> g_E=1.05827M$, the prograde and retrograde photon orbits exist for $g\leq 1.2183407M$ and lead to a closed photon ring structure. For $a=0.1M$, $g_E=1.05231M$ and $g_p=1.17566M$ (cf. Fig.~\ref{fig:NoBHShadow}). The sizes of the black hole shadows decrease with both $a$ and $g$, and constraints on the parameters are derived from the Sgr A* black hole shadow observation from the EHT. 

\subsection{Ghosh-Culetu Spacetime}\label{sec:gc}
Bardeen and Hayward regular black holes have an asymptotically de-Sitter core. The next regular model \citep{Ghosh:2014pba,Culetu:2014lca,Singh:2022xgi}, which we hereafter refer to as Ghosh-Culetu black holes, is a novel class of regular black hole with an asymptotic Minkowski core \citep{Simpson:2019mud}. While these regular models share many features with Bardeen and Hayward black holes, there are also notable differences, especially at the deep core \citep{Simpson:2019mud}. This solution was found by Ghosh  \citep{Ghosh:2014pba} generalizing the
spherically symmetric regular (i.e., singularity-free) black hole
solution \citep{Culetu:2014lca} to the rotating case. The mass function of rotating non-singular black hole reads  \citep{Ghosh:2014pba}
\begin{eqnarray}
m(r)=Me^{-g^2/2Mr},
\end{eqnarray}
where $g$ is the NED charge. 
The  Ghosh-Culetu black hole is a generalization of the Kerr black hole that goes over asymptotically as the Kerr-Newman black hole and, in the limit, $g\to0$ as the Kerr black hole. The size of a shadow reduces as the parameter $g$ increases, and the shadow becomes more distorted as we increase the parameter $g$. Again, the bound $\Delta C \leq 0.1$ merely constrains the black hole parameter space, and theoretically allowed values of $g$, that are consistent with the observed asymmetry of M87*, is  $g \leq 1.2130M$. On the other hand, the observed angular diameter $\theta_{sh}$ of the M87* black hole, within the $1 \sigma$ confidence level, requires $g \leq 0.30461M$ for the Ghosh-Culetu black holes \citep{Kumar:2020yem}. 

Although, for a fixed value of $a$, the black hole horizons disappear for $g>g_E$, the prograde and retrograde photon orbits exist for $g\leq g_p$ and lead to a closed photon ring structure. The values of $g_E$ and $g_p$ are summarized in Table~\ref{Table1}.  
For $a=0$, the $g_E=1.21306M$ and $g_p=1.26453M$, whereas for $a=0.1M$, $g_E=1.202M$ and $g_p=1.2291M$. Black hole shadow size decreases with both $a$ and $g$, and constraints on parameters are derived from the Sgr A* black hole shadow observation from the EHT (cf. Fig. \ref{fig:BHShadow} ). 

\subsection{Simpson-Visser Spacetime}\label{sec:sv}
Simpson and Visser proposed an interesting spherically symmetric regular black hole spacetime     \citep{Simpson:2019cer} with an additional parameter $g$ that characterize the least modification of the Schwarzschild black hole spacetime. Simpson-Visser spacetime  interpolates between the Schwarzschild black hole and the Morris--Thorne traversable wormhole and is regular everywhere. It has been shown that the Simpson-Visser  metric solves the Einstein field equations sourced by a phantom scalar field with minimally coupled to the nonlinear electrodynamics \citep{Bronnikov:2021uta}.

\begin{figure*}[ht!]
\begin{tabular}{c c}
	\includegraphics[scale=0.75]{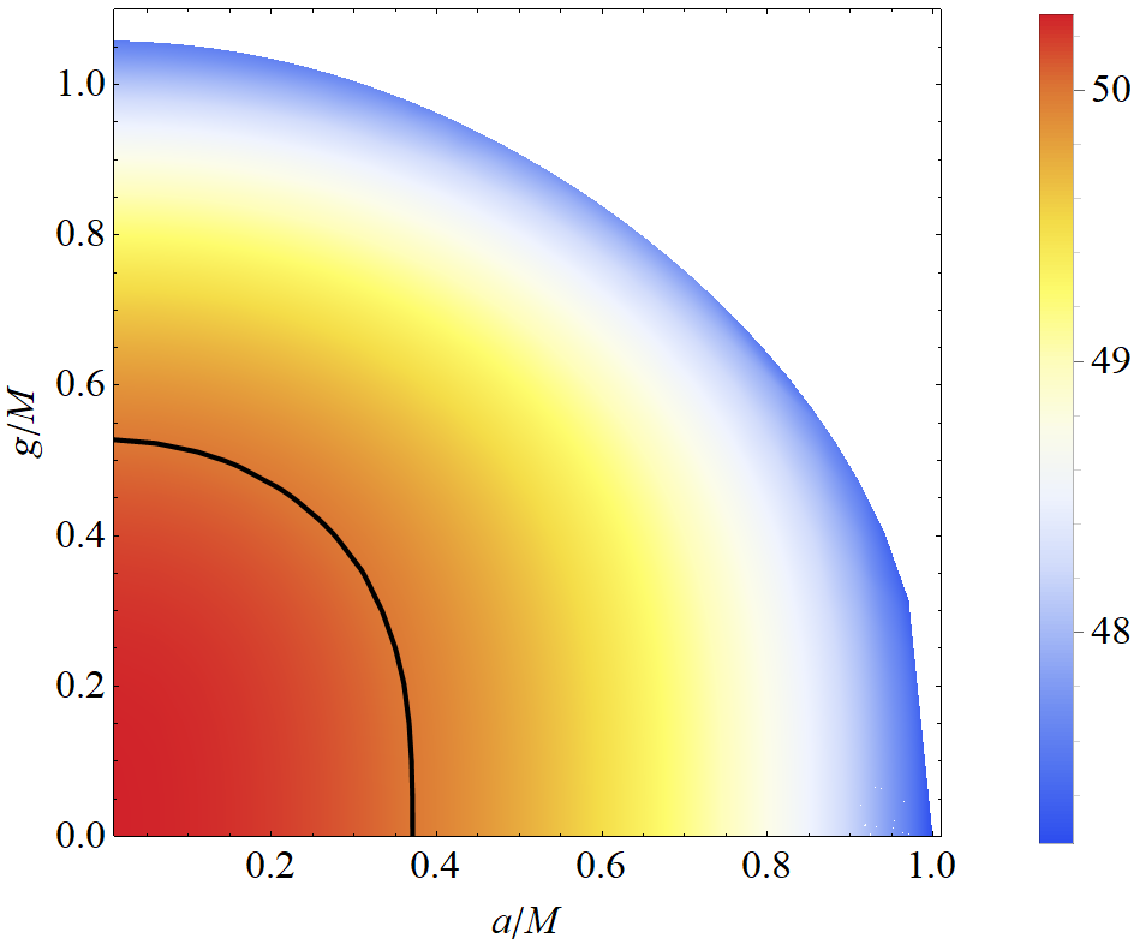}&
	\includegraphics[scale=0.75]{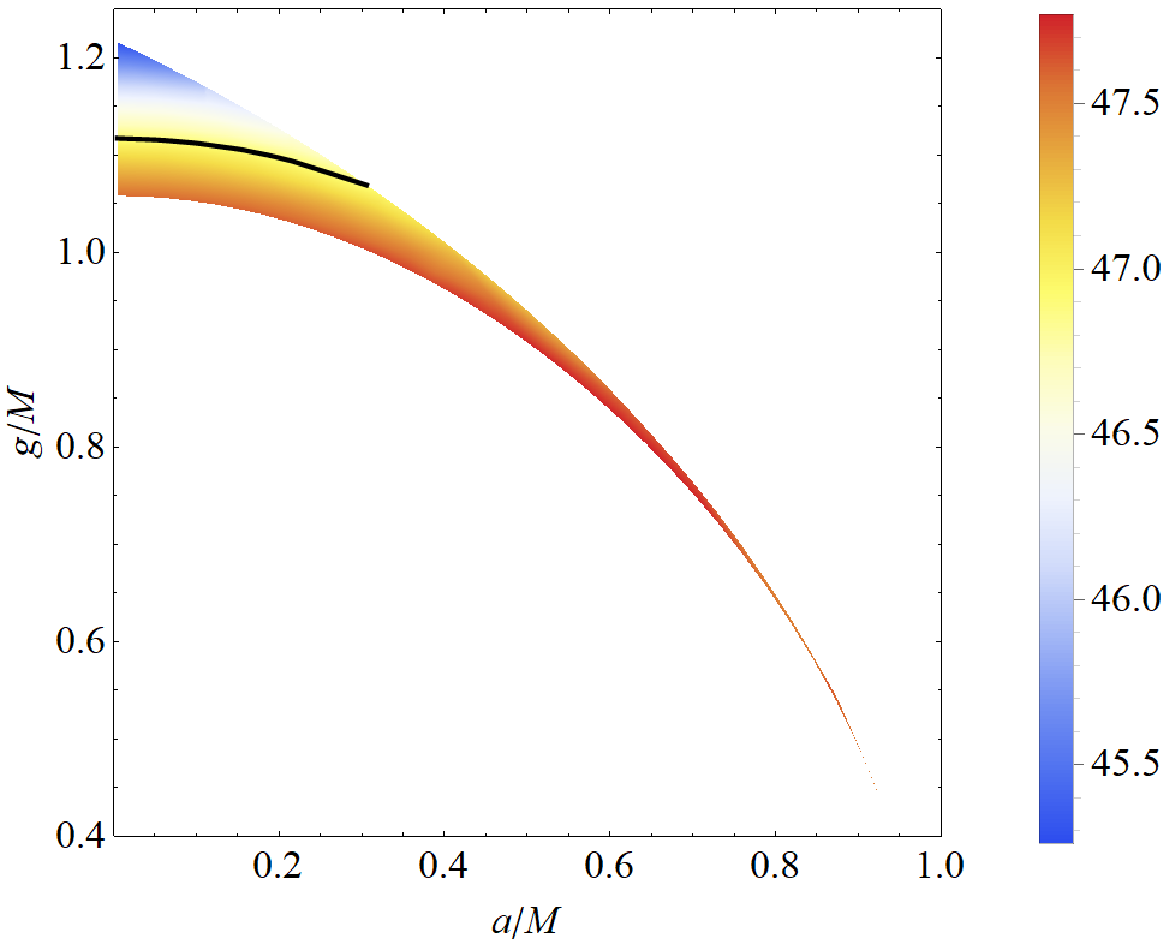}
\end{tabular}	
	\caption{(Left:) Hayward black holes' shadow angular diameter $\theta_{sh}$, in  $\mu$as, as a function of ($a, g$). The black line is for $\theta_{sh}=50\mu$as corresponding to the Sgr A* black hole shadow. Hayward black holes that have parameters outside of the black line produce a shadow that is consistent with the Sgr A* shadow size. (Right:) Hayward no-horizon spacetime shadow angular diameter $\theta_{sh}$, in units of $\mu$as, as a function of ($a, g$). The black line is for $\theta_{sh}=46.9\mu$as corresponding to the Sgr A* black hole shadow, such that the region under the black line satisfies the Sgr A* shadow size.}\label{fig:HayAng}
\begin{tabular}{c c}
	\includegraphics[scale=0.75]{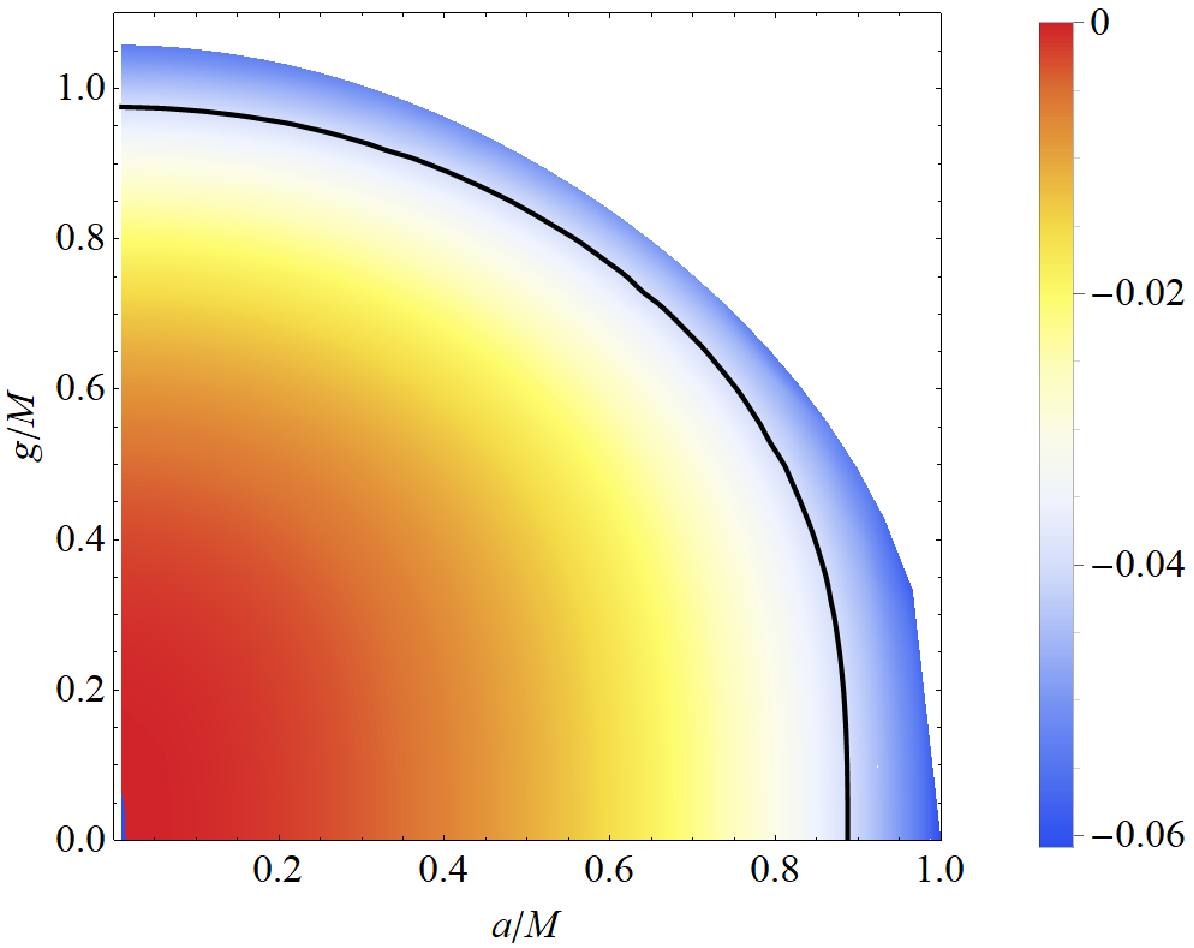}&
	\includegraphics[scale=0.75]{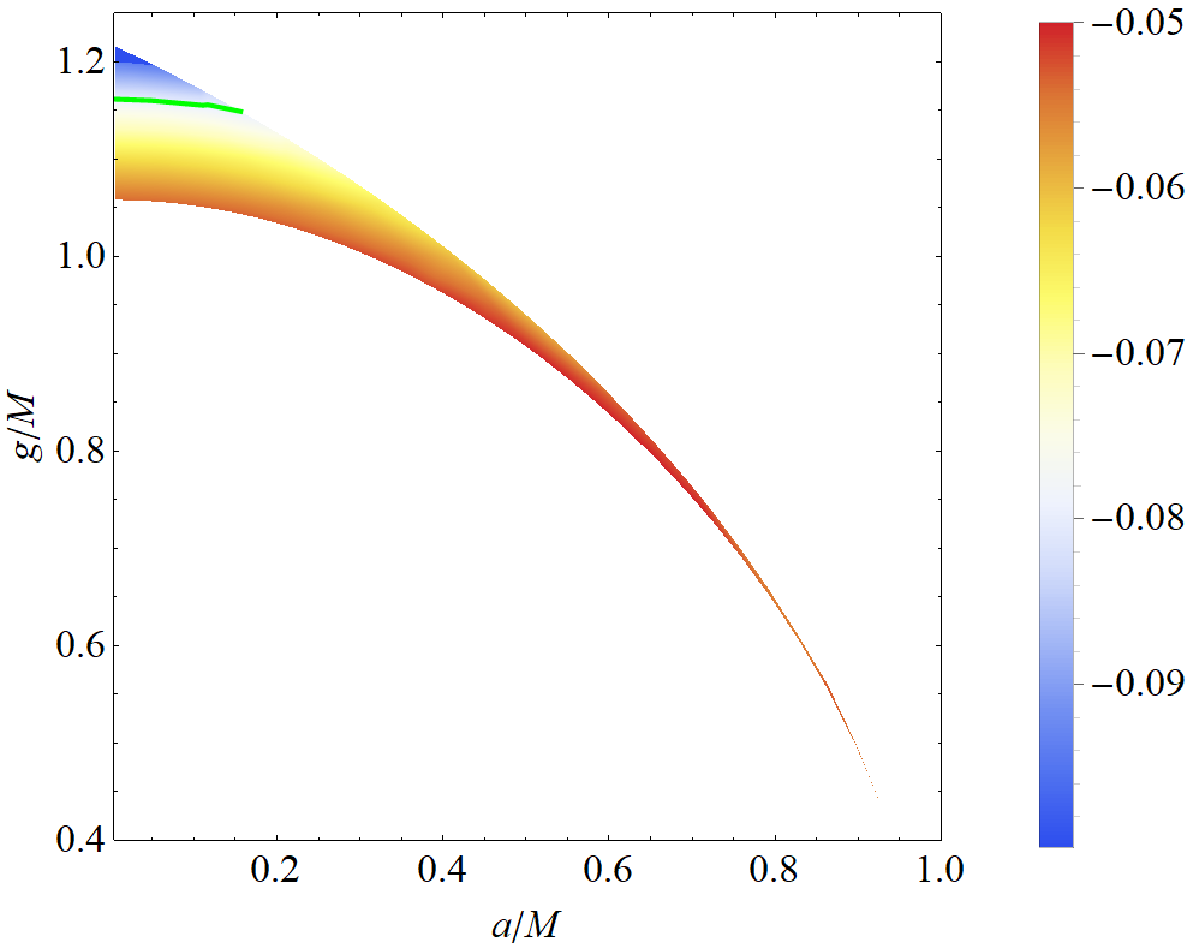}
\end{tabular}	
	\caption{Hayward black holes (left) and no-horizon spacetime (right) shadow angular diameter deviation from that of a Schwarzschild black hole as a function of ($a, g$). The black and green lines are, respectively, for the  $\delta=-0.04$ (Keck)  and $\delta=-0.08$ (VLTI) corresponding to the Sgr A* black hole shadow bounds. }\label{fig:HayDelta}
\end{figure*}
\begin{figure*}
\begin{tabular}{c c}
	\includegraphics[scale=0.75]{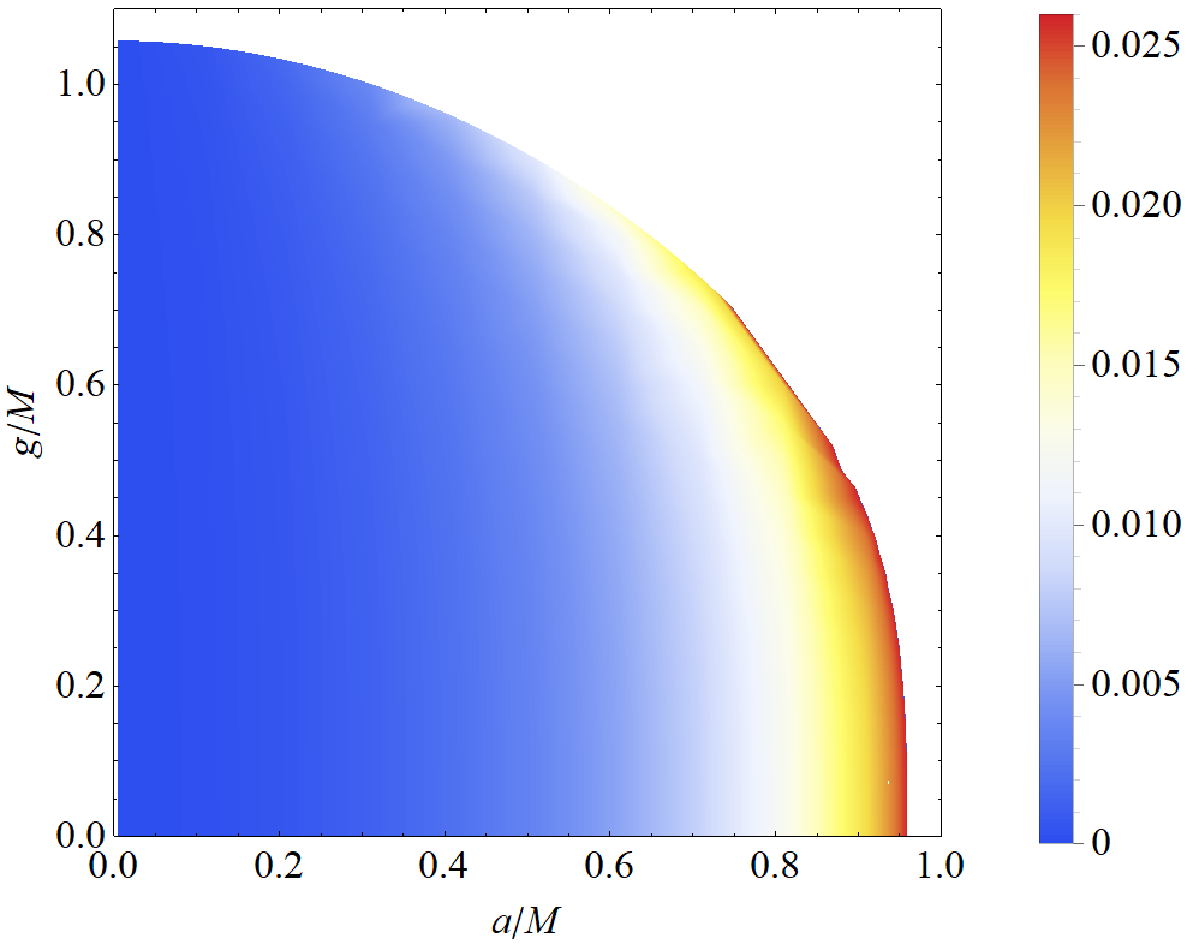}&
	\includegraphics[scale=0.75]{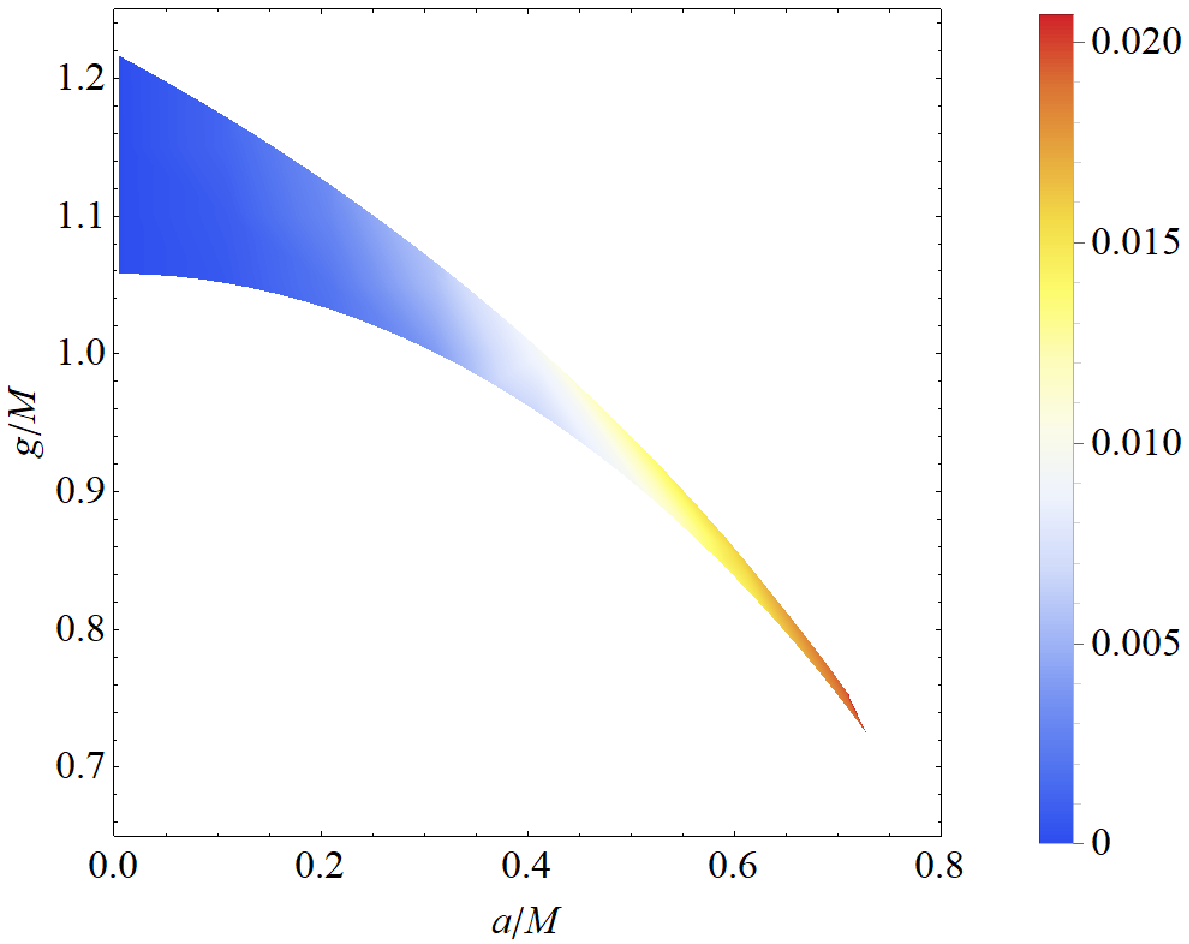}
\end{tabular}	
	\caption{Hayward black holes (left) and no-horizon spacetime (right) shadows circularity deviation observable $\Delta C$ as a function of ($a, g$). }\label{fig:HayAsymm}
\end{figure*}

The rotating Simpson-Visser black holes metric is constructed as a modification from the Kerr black hole metric in Refs.~\citep{Shaikh:2021yux,Mazza:2021rgq}, and is given by the (\ref{rotmetric}) with mass function 
\begin{eqnarray}
m(r)=M \sqrt{1+ \frac{g^2}{r^2}}.
\end{eqnarray}
$\Sigma = r^2+g^2 +a^2\cos^2\theta$ and  $\Delta = r^2+g^2+a^2-2 m(r) r$ are different from earlier three black holes. It is important to note here that the radial coordinate $r$ is not the same as the areal radius, which is $\sqrt{r^2+g^2}$. The Simpson-Visser spacetime has an interesting feature, namely that the areal radii of the event horizon, the photon sphere, and the shadow are completely independent of parameter $g$, but, the proper distance between these surfaces is $g-$dependent \citep{Lima:2021las}. Several studies have been done on the observational features of Simpson-Visser black hole spacetime, namely gravitational lensing \citep{Tsukamoto:2020bjm,Ghosh:2022mka}, the black hole shadow \citep{Lima:2021las,Shaikh:2021yux,Jafarzade:2021umv}, the optical appearance of a thin accretion disk \citep{Bambhaniya:2021ugr,Guerrero:2021ues}, quasi-periodic oscillations \citep{Jiang:2021ajk}. Although, for $a=0$, the black hole horizons disappear for $g> g_E=2.0M$, the prograde and retrograde photon orbits exist for $g\leq 3.0M$ and lead to a closed photon ring structure. For $a=0.1M$, $g_E=1.995M$ and $g_p=2.9083M$. Interestingly, the black hole shadow size decreases only with $a$, and it remains completely independent of $g$. We assume that the rotating regular spacetimes defined by the metric (\ref{rotmetric}) describe supermassive compact objects Sgr A* focusing on four well-motivated different spacetimes, and determine whether their images are compatible with those observed with EHT, and constraints the deviation parameter. 

\section{Observational constraints from the EHT results of  Sgr A*}\label{sect4}

\begin{figure*}
\begin{tabular}{c c}
	\includegraphics[scale=0.75]{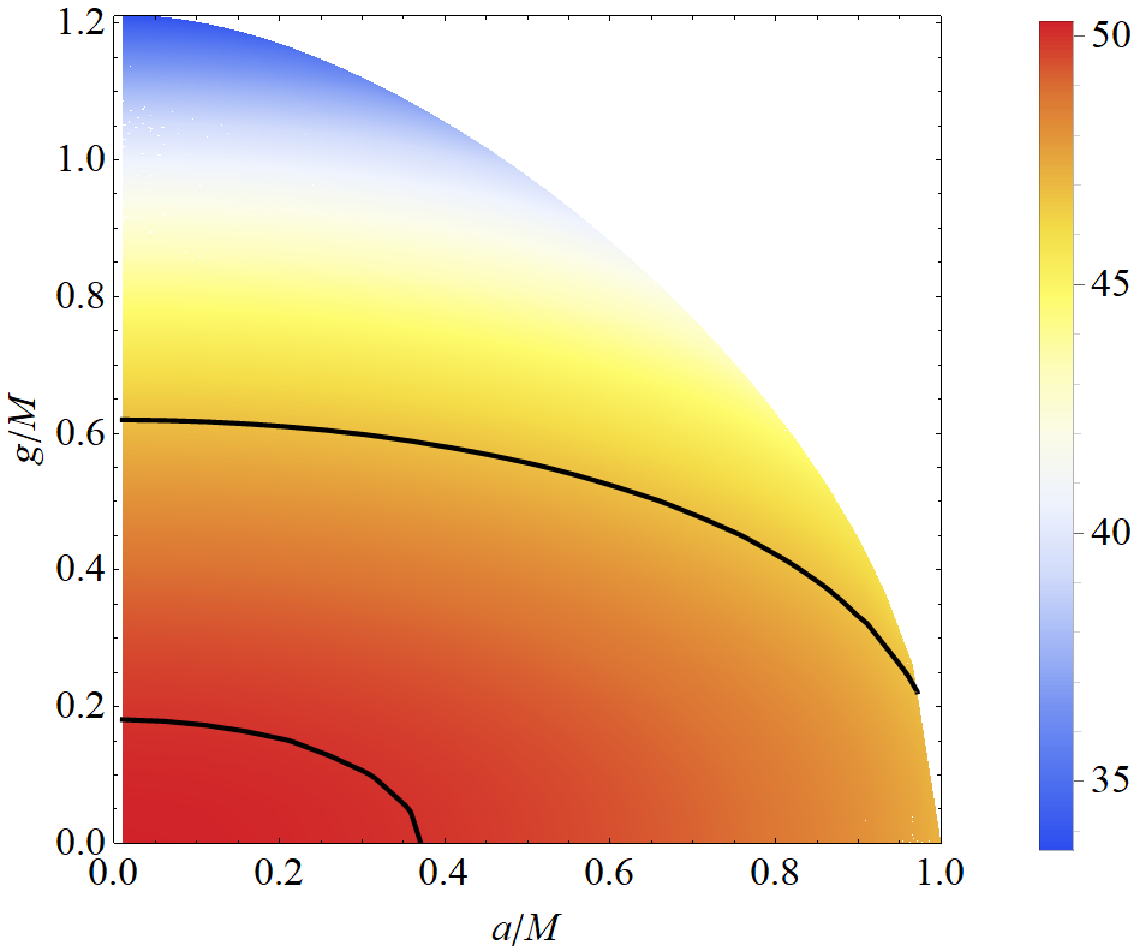}&
	\includegraphics[scale=0.75]{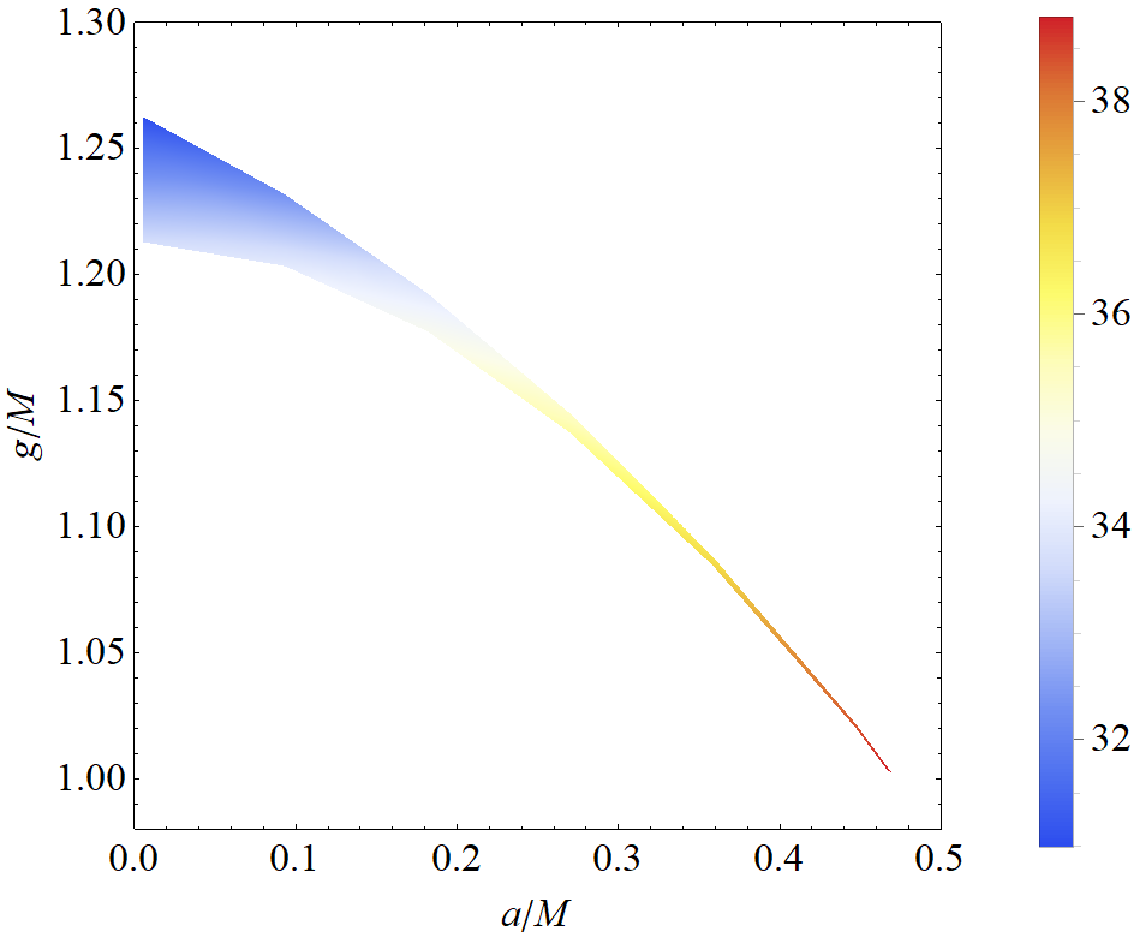}
\end{tabular}	
	\caption{(Left:) Ghosh-Culetu black holes' shadow angular diameter $\theta_{sh}$, in units of $\mu$as, as a function of ($a, g$). Black lines are for $46.9\mu$as and $50\mu$as corresponding to the Sgr A* black hole shadow bounds. Ghosh-Culetu black holes that have parameters between the two black lines produce a shadow that is consistent with the magnitude of the Sgr A* shadow. (Right:) Bardeen no-horizon spacetime shadow angular diameter $\theta_{sh}$, in units of $\mu$as, as a function of ($a, g$).}\label{fig:NSAng}
\begin{tabular}{c c}
	\includegraphics[scale=0.75]{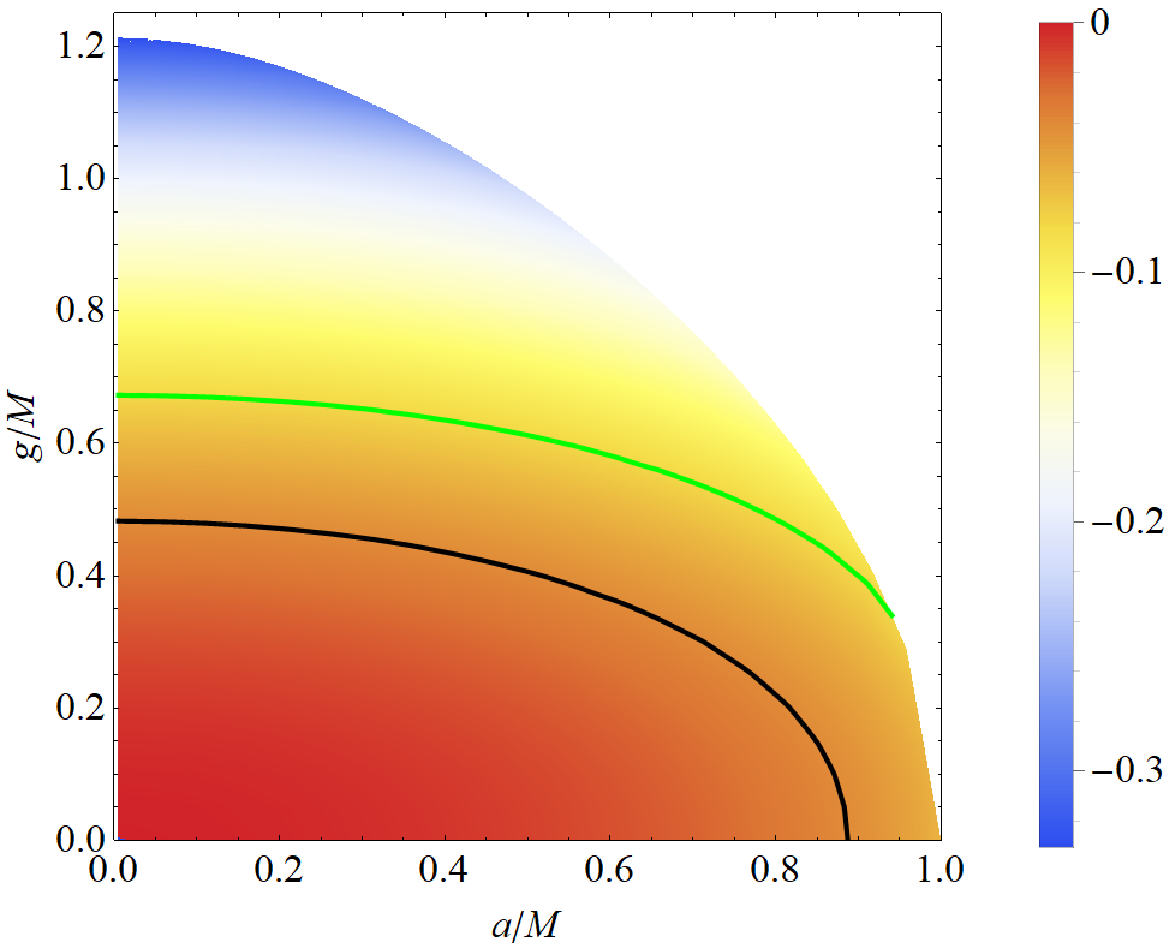}&
	\includegraphics[scale=0.75]{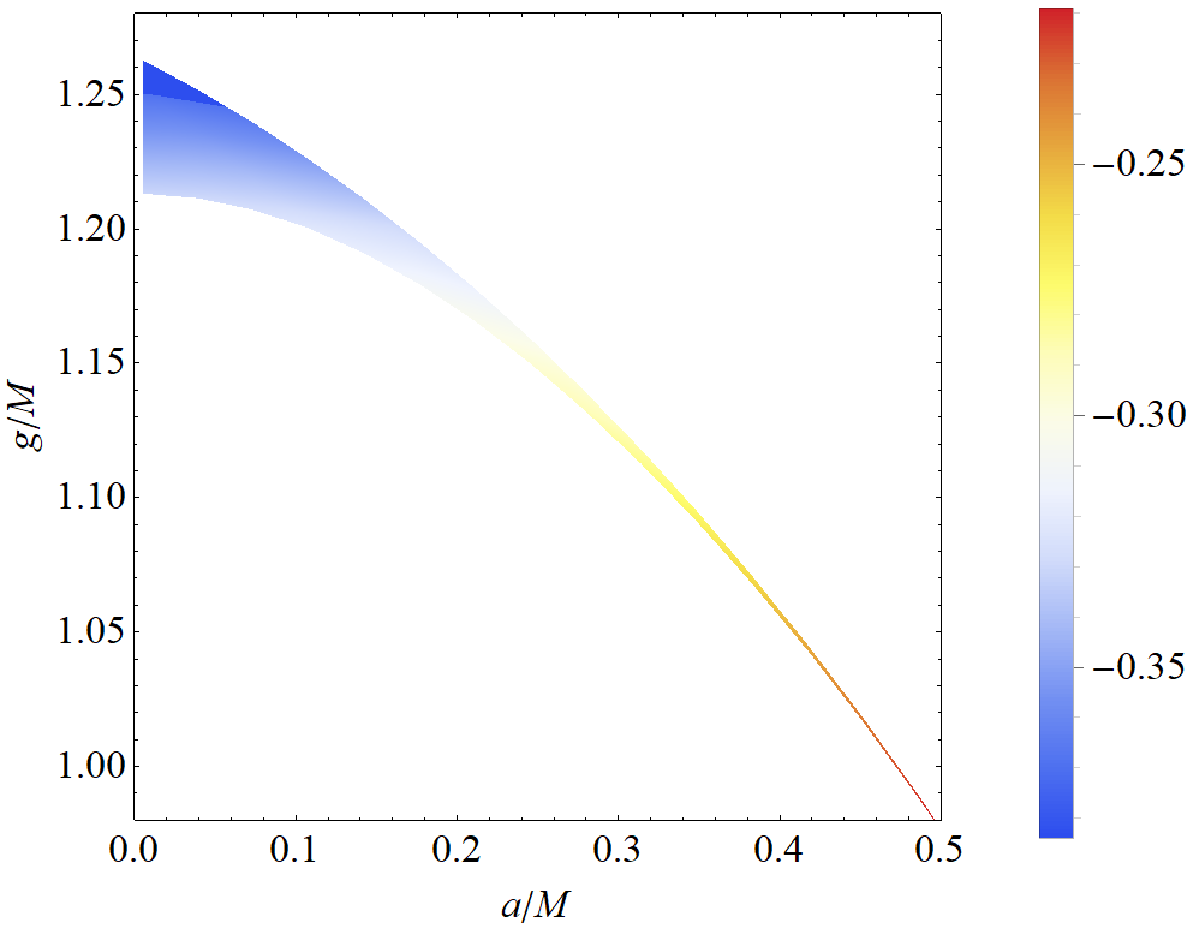}	
\end{tabular}	
	\caption{Ghosh-Culetu black holes (left) and no-horizon spacetime (right) shadow angular diameter deviation from that of a Schwarzschild black hole as a function of ($a, g$). The black and green lines are, respectively, for the  $\delta=-0.04$ (Keck)  and $\delta=-0.08$ (VLTI) corresponding to the Sgr A* black hole shadow bounds. }\label{fig:NSDelta}
\end{figure*}
\begin{figure*}
\begin{tabular}{c c}
	\includegraphics[scale=0.75]{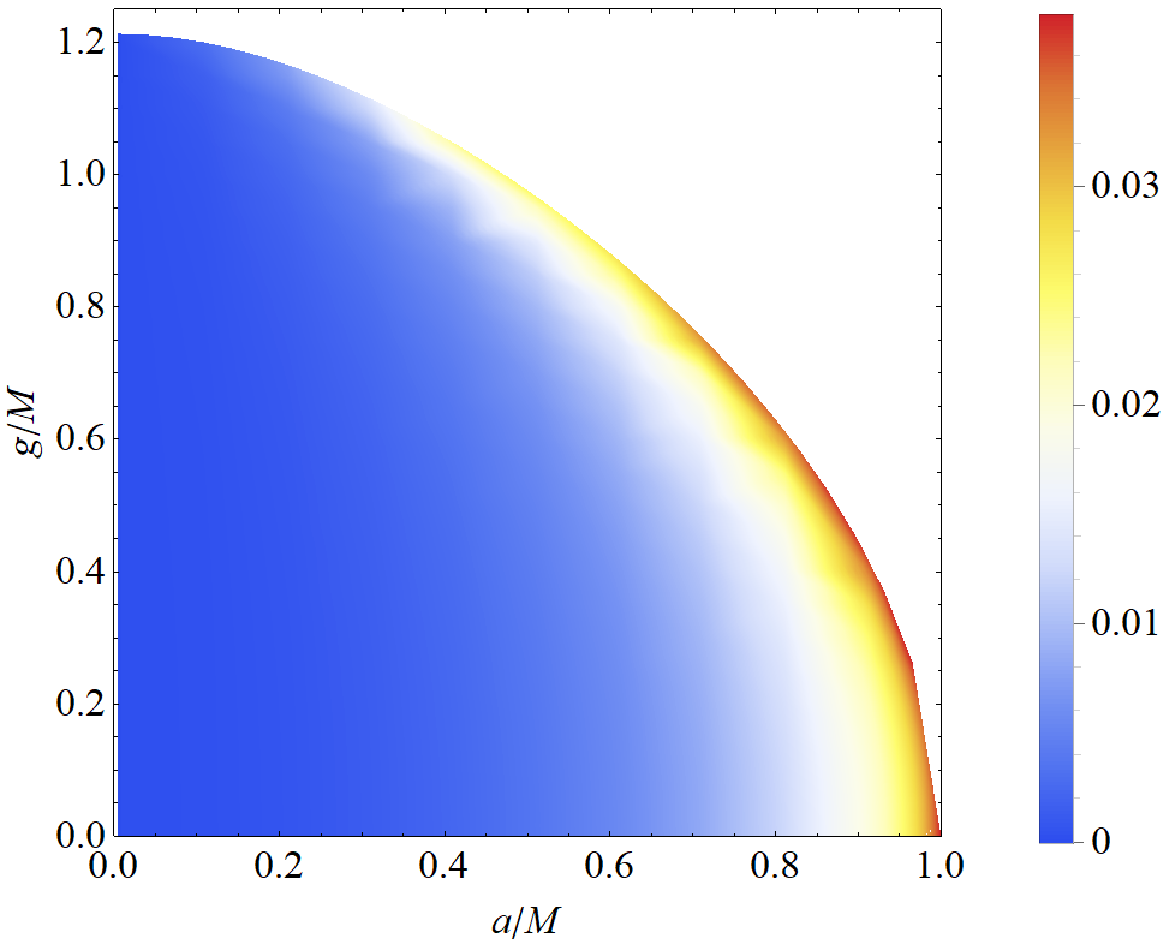}&
	\includegraphics[scale=0.75]{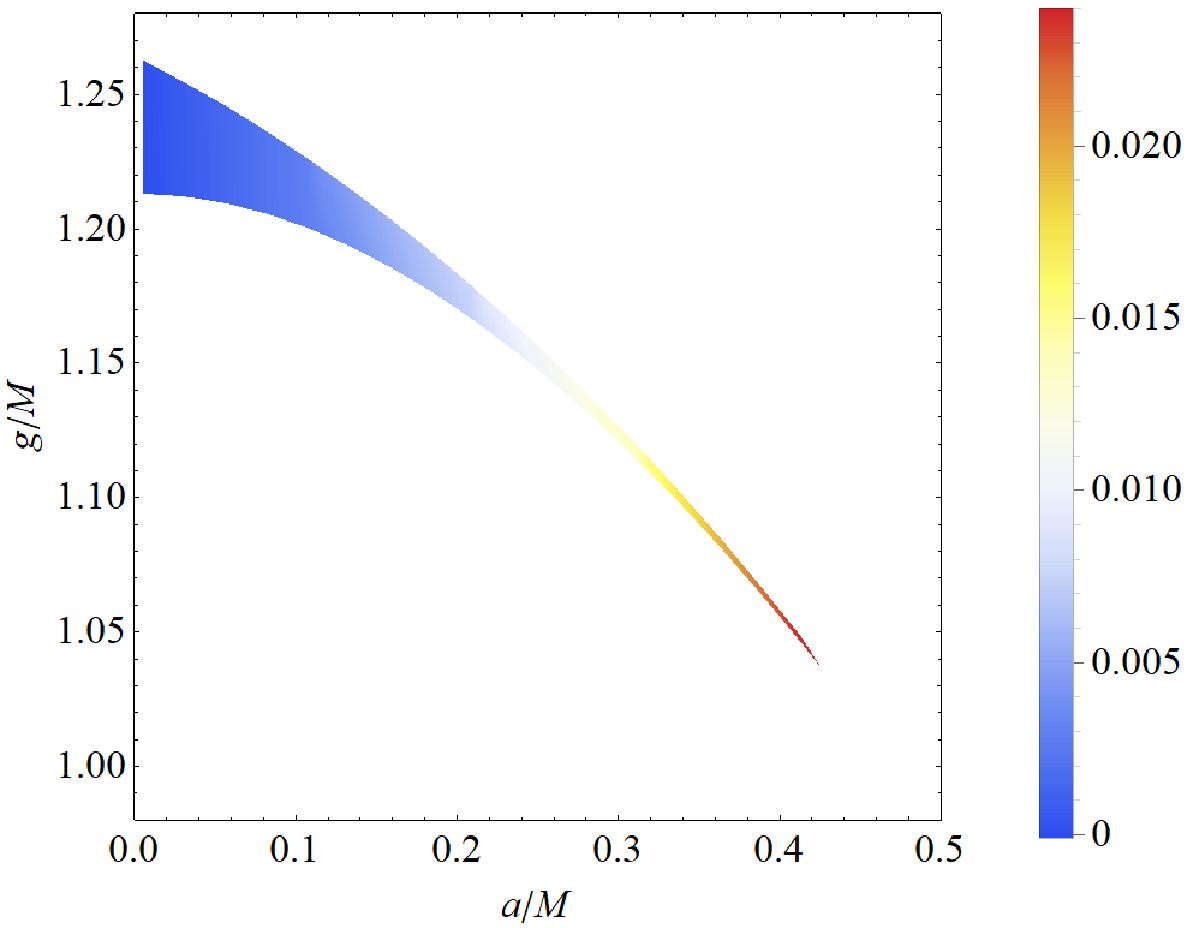}
\end{tabular}	
	\caption{Ghosh-Culetu black holes (left) and no-horizon spacetime (right) shadows  circularity deviation observable $\Delta C$ as a function of ($a, g$). }\label{fig:NSAsymm}
\end{figure*}

\begin{figure*}
\begin{tabular}{c c}
	\includegraphics[scale=0.75]{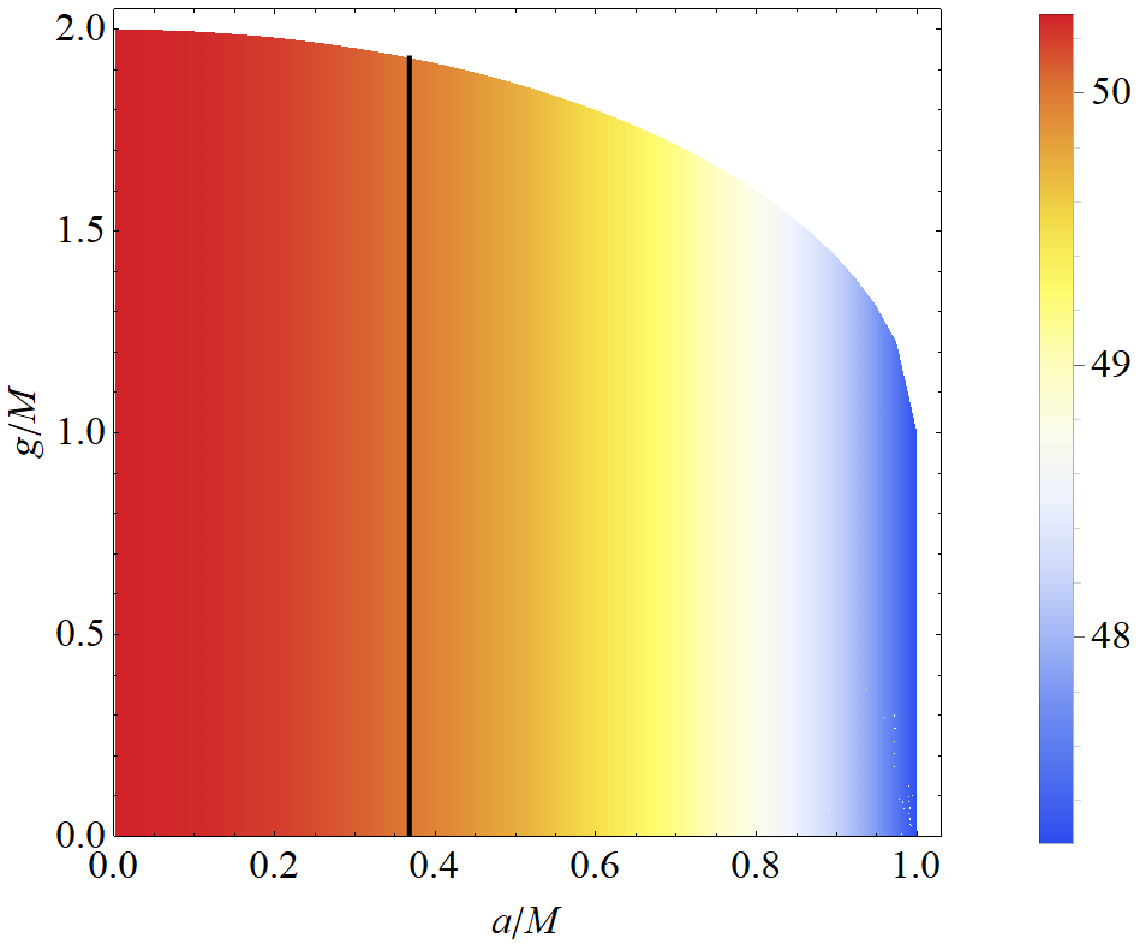}&
	\includegraphics[scale=0.75]{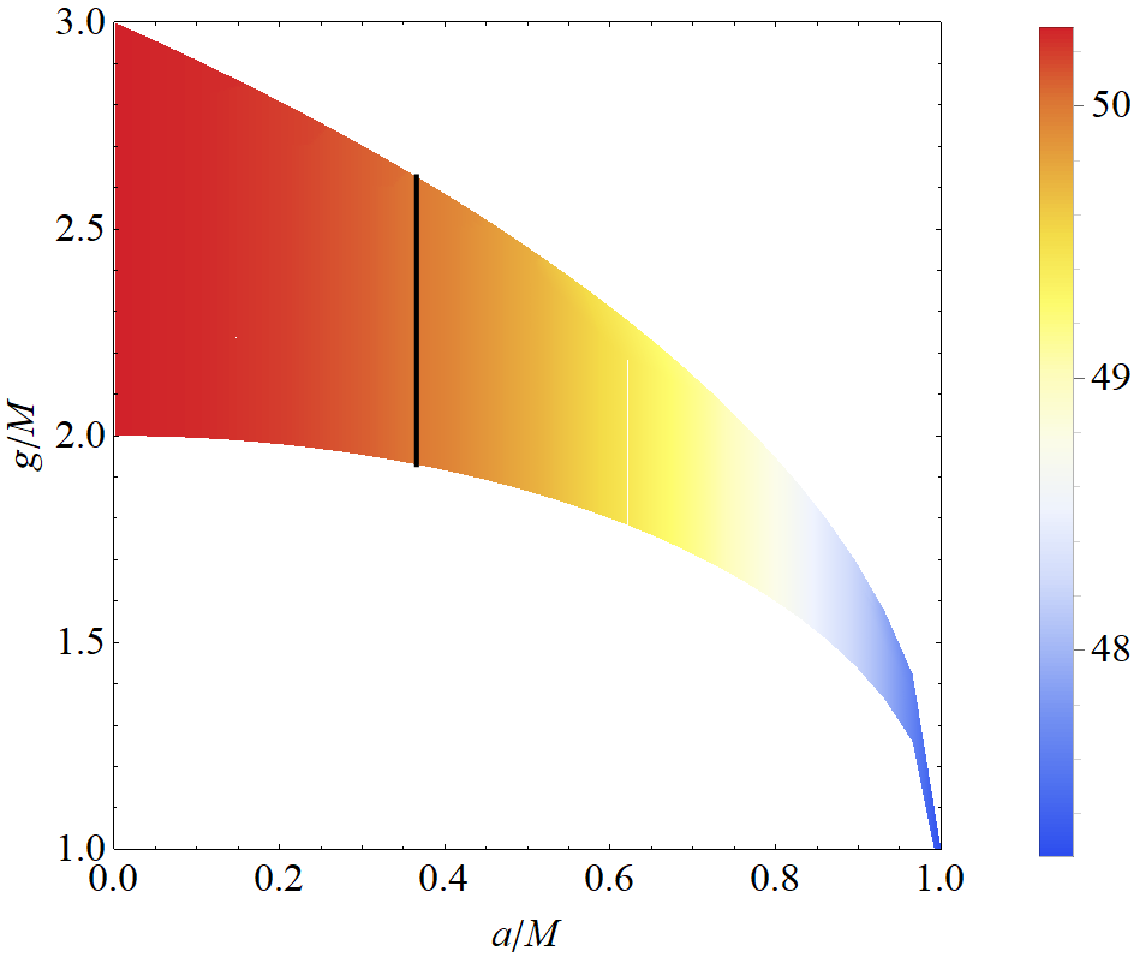}
\end{tabular}	
	\caption{Simpson-Visser black holes (left) and no-horizon spacetime (right) shadow angular diameter $\theta_{sh}$, in units of $\mu$as, as a function of ($a, g$). Black line is for $46.9\mu$as corresponding to the Sgr A* black hole shadow bounds. }\label{fig:SVAng}
\begin{tabular}{c c}
	\includegraphics[scale=0.75]{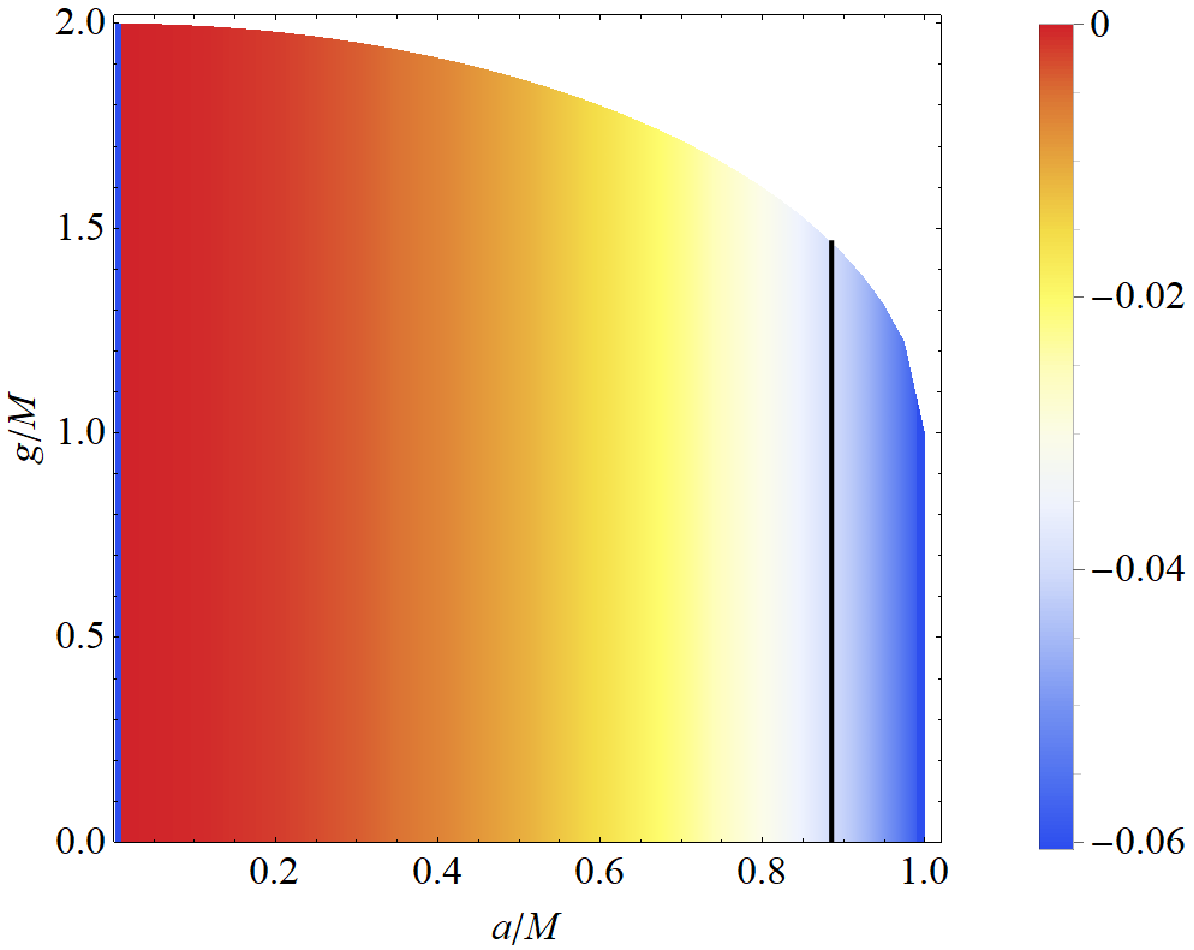}&
	\includegraphics[scale=0.75]{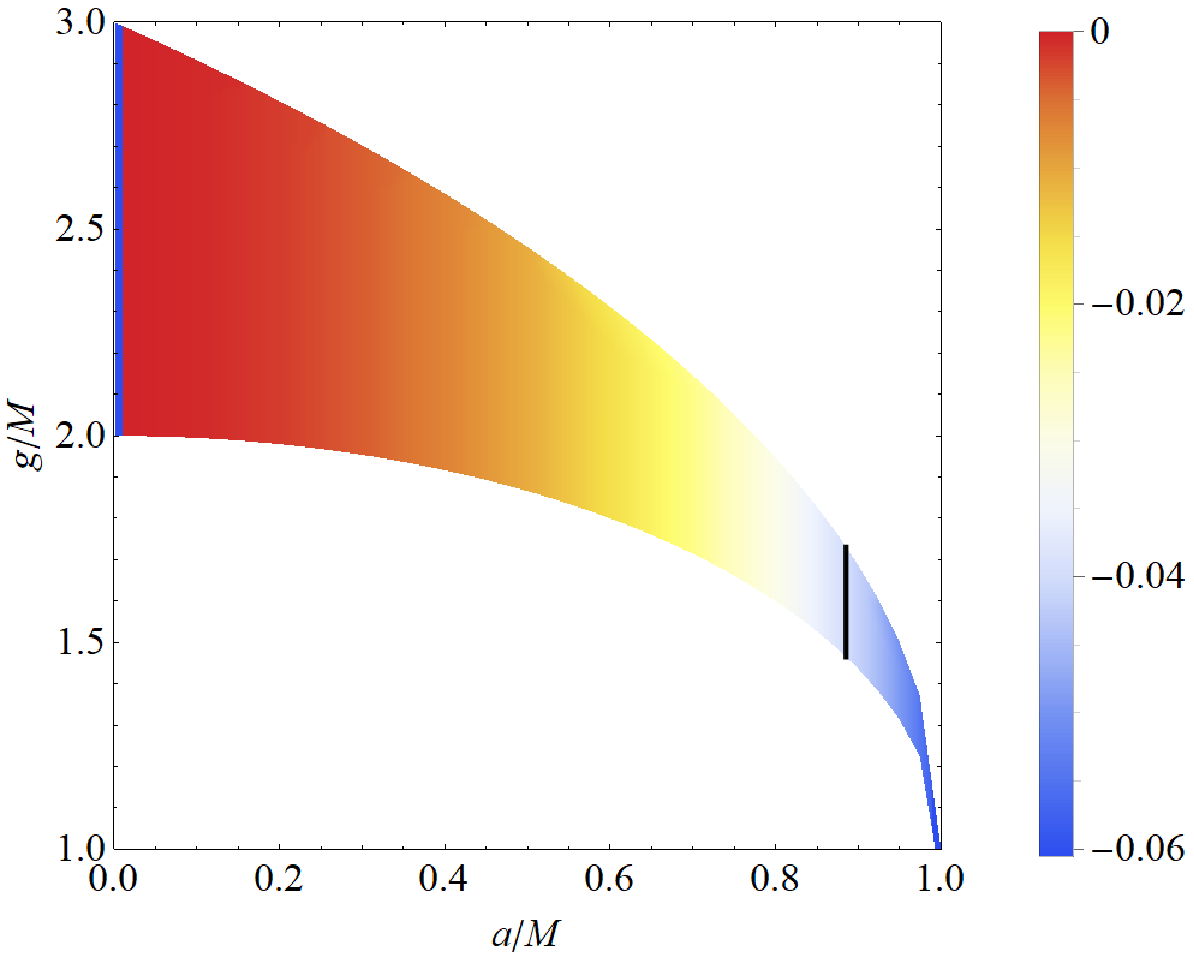}	
\end{tabular}	
	\caption{Simpson-Visser black holes (left) and no-horizon spacetime (right) shadows angular diameter deviation from that of a Schwarzschild black hole as a function of ($a, g$). The black and green lines are, respectively, for the  $\delta=-0.04$ (Keck)  and $\delta=-0.08$ (VLTI) corresponding to the Sgr A* black hole shadow bounds. }\label{fig:SVDelta}
\end{figure*}
\begin{figure*}
\begin{tabular}{c c}
	\includegraphics[scale=0.75]{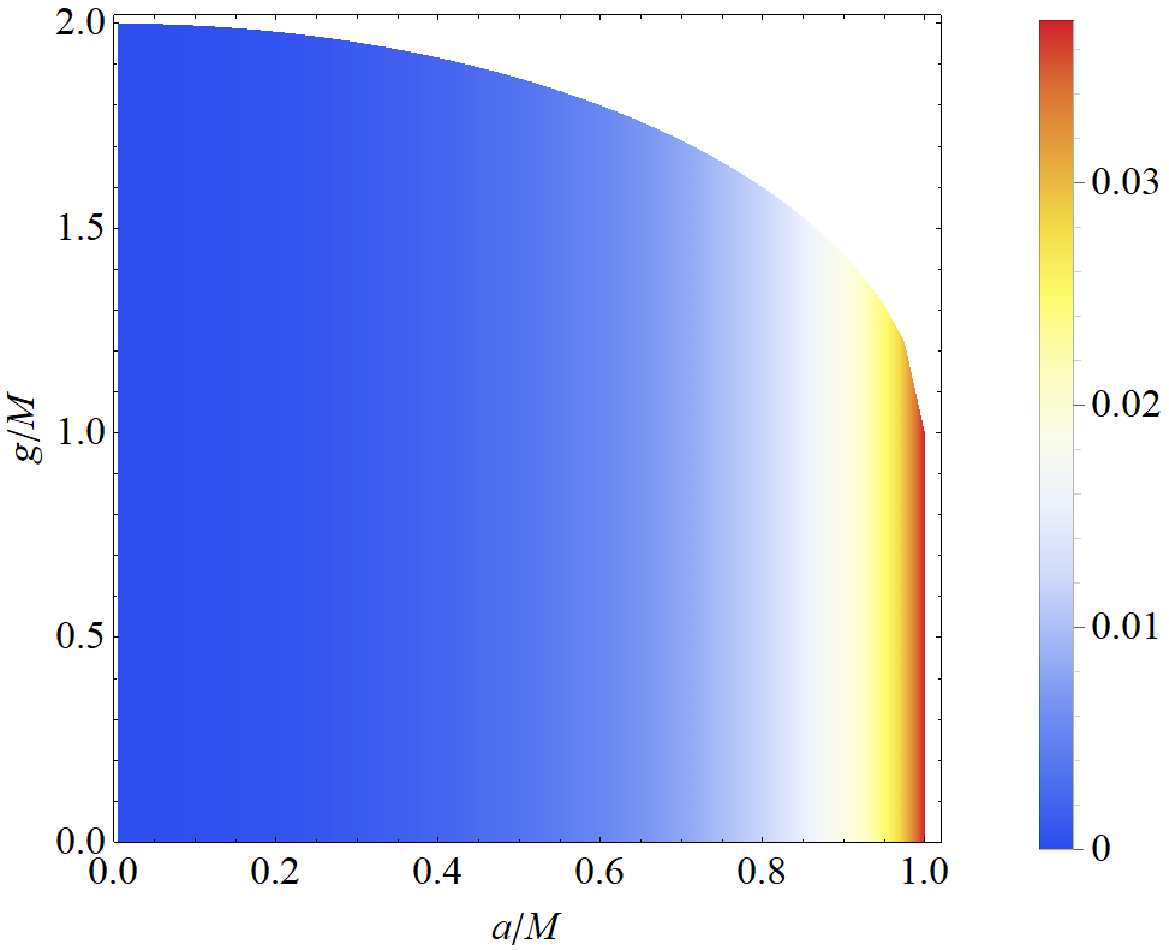}&
	\includegraphics[scale=0.75]{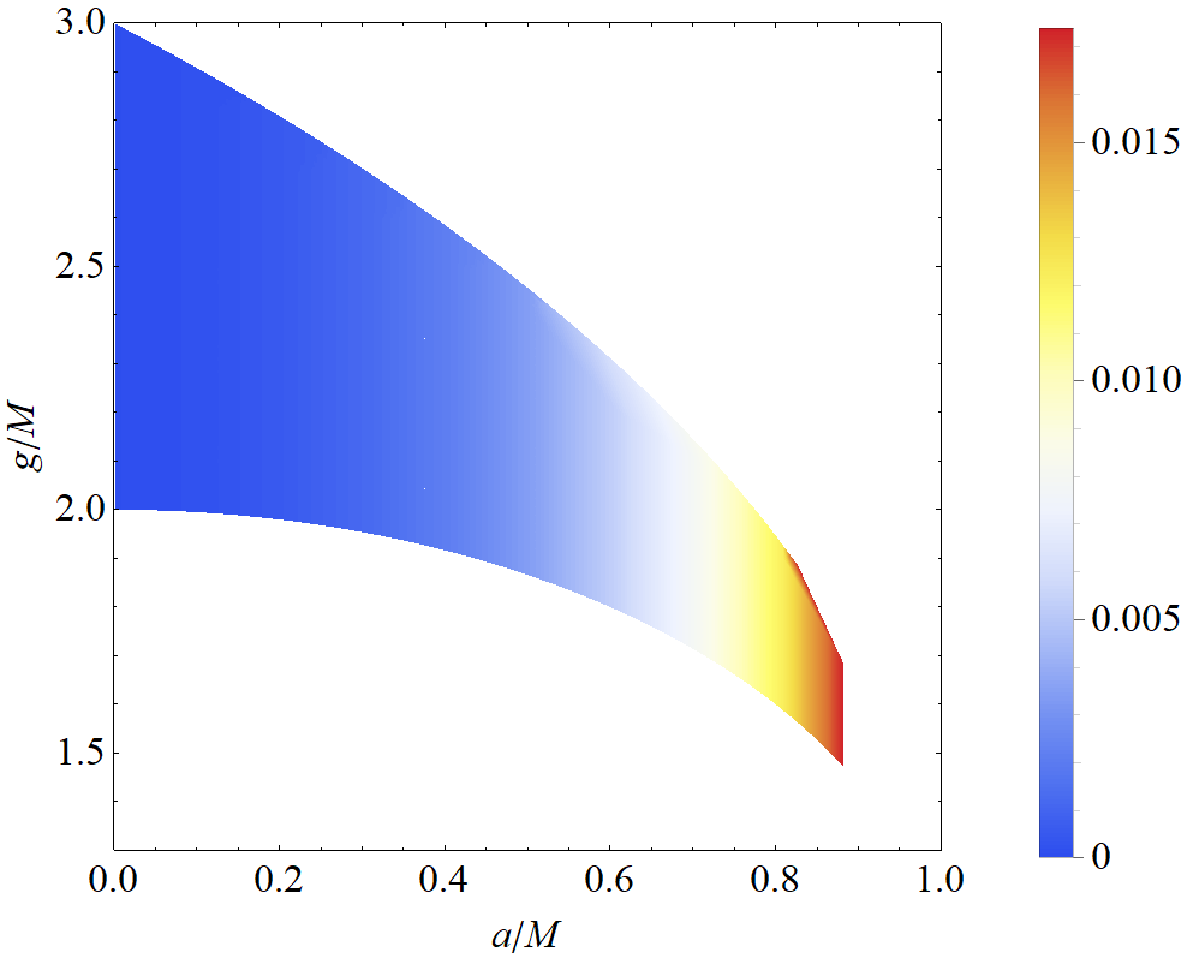}
\end{tabular}	
	\caption{Simpson-Visser black holes (left) and no-horizon spacetime (right) shadows  circularity deviation observable $\Delta C$ as a function of ($a, g$). }\label{fig:SVAsymm}
\end{figure*}

The EHT collaboration has released the shadow results for the Sgr A* black hole based on the VLBI observational campaign at 1.3mm wavelength conducted in 2017 \citep{EventHorizonTelescope:2022exc,EventHorizonTelescope:2022urf,EventHorizonTelescope:2022vjs,EventHorizonTelescope:2022wok,EventHorizonTelescope:2022xnr, EventHorizonTelescope:2022xqj}. Even though the mass and distance of Sgr A* are already well-known from the resolved individual star orbits, the horizon scaled shadow measurements also provided similar estimates \citep{Do:2019txf,abuter2019geometric}. Various imaging and modeling techniques were used to construct the shadow images, which are all remarkably consistent in having a central brightness depression encircled by a bright emission ring \citep{EventHorizonTelescope:2022exc,EventHorizonTelescope:2022xqj}. The emission ring corresponds to the lensed position of the surrounding accretion disk inner edge and thus different from the characteristics photon ring (shadow boundary), which preferentially falls inside the emission ring. The photon ring solely depends on the black hole spacetime, whereas the emission ring additionally depends on the accretion model. Thus, the photon ring can be used as a tool to test the black hole matrices. In contrast to the M87* black hole, for which the EHT only provided an estimate of the emission ring diameter \citep{Akiyama:2019cqa}, for Sgr A*, the EHT not only measured the emission ring angular diameter $\theta_d=(51.8\pm 2.3)\mu$as but also estimated the shadow diameter $\theta_{sh}=(48.7\pm 7)\mu$as with the priors $M=4.0^{+1.1}_{-0.6} \times 10^6 M_{o}$ and $D_{LS}=8.15\pm 0.15$ kpc \citep{EventHorizonTelescope:2022xqj}. For instance, EHT employed three independent imaging algorithms, \texttt{ eht-imaging}, \texttt{SIMLI}, and \texttt{DIFMAP}, to determine the Sgr A* shadow morphology. The most likely averaged measured value of the shadow angular diameter from these three algorithms is in the range $\theta_{sh}\in (46.9-50)\,\mu$as, and the $1\sigma$ credible interval is $41.7-55.6\,\mu$as \citep{EventHorizonTelescope:2022xqj}. Here, we consider the Sgr A* black hole as the rotating regular black hole model and use the Sgr A* shadow results to test the viability of these models to explain the astrophysical black hole spacetimes. For this purpose, we will use three observables, the shadow angular diameter $\theta_{sh}$, the deviation of the shadow size from that of a Schwarzschild black hole $\delta$, and the circular asymmetry $\Delta C$. The shadow angular diameter, defined in terms of the shadow cone opening angle, reads as
\begin{equation}
\theta_{sh}=\frac{2}{D_{LS}}\sqrt{\frac{A}{\pi}},
\end{equation}    
where $A$ is the black hole shadow area \citep{Kumar:2018ple,Abdujabbarov:2015xqa}
\begin{equation}
A=2\int{Y(r_p) dX(r_p)}=2\int_{r_p^{-}}^{r_p^+}\left( Y(r_p) \frac{dX(r_p)}{dr_p}\right)dr_p.\label{Area}
\end{equation} 
Furthermore, EHT used the observable $\delta$ to quantify the deviation between the inferred shadow diameter and that of a Schwarzschild black hole as follows \citep{EventHorizonTelescope:2022urf,EventHorizonTelescope:2022xqj}
\begin{equation}
\delta=\frac{\theta_{sh}}{\theta_{sh, Sch}}-1.
\end{equation}
Clearly, $\delta$ takes positive (negative) values if the black hole shadow size is greater (smaller) than the Schwarzschild black hole of the same mass. 
EHT used the two separate priors for the Sgr A* angular size from the Keck and Very Large Telescope Interferometer (VLTI) observations and the three independent imaging models to estimate the bounds on the fraction deviation observable $\delta$ \citep{EventHorizonTelescope:2022exc, EventHorizonTelescope:2022xqj}
\begin{align}
\delta= \begin{dcases*} -0.08^{+0.09}_{-0.09}\;\;\;\;\; & \text{VLTI}\\
		  -0.04^{+0.09}_{-0.10}\;\;\;\;\; &\text{Keck}	 
		\end{dcases*} 
\end{align}
In addition, we define the dimensionless circularity deviation observable as a measure of the distortion of the shadow from a perfect circle described by polar coordinates ($R(\varphi), \varphi$) \citep{Johannsen:2010ru,Johannsen:2015qca, Kumar:2020yem}
\begin{equation}
\Delta C=\frac{1}{\bar{R}}\sqrt{\frac{1}{\pi}\int_0^{2\pi}\left(R(\varphi)-\bar{R}\right)^2d\varphi},
\end{equation}
where $\bar{R}$ is the average shadow radius defined as \citep{Johannsen:2010ru}
\begin{equation}
\bar{R}=\frac{1}{2\pi}\int_{0}^{2\pi} R(\varphi) d\varphi.
\end{equation}
The deviation of the black hole shadow from a circle strongly depends on the spin parameter. However, the EHT 2017 observations could not estimate the Sgr A* shadow circularity deviation nor the Sgr A* black hole spin \citep{EventHorizonTelescope:2022xqj}. Therefore, to constrain the rotating regular black hole parameters ($a, g$), we will use the observed bounds of $\theta_{sh}$ and $\delta$ for the Sgr A* shadow.

We calculated and depicted the Bardeen black holes and corresponding no-horizon spacetime shadow angular diameter $\theta_{sh}$ in Fig.~\ref{fig:BarAng}, the shadow angular diameter deviation $\delta$ in Fig.~\ref{fig:BarDelta}, and the circularity deviation $\Delta C$ in Fig.~\ref{fig:BarAsymm}. As shown in Fig.~\ref{fig:BarAng}, for $a=0$, Bardeen black holes with $0.18M\leq g \leq  0.595M$ cast shadows with an angular diameters $\theta_{sh}\in (46.9, 50)\mu$as corresponding to the observed Sgr A* shadow angular diameter, whereas no-horizon spacetimes with $0.7698M\leq g \leq 0.83794M$ satisfy the $1\sigma$ bound $41.7-55.6\,\mu$as. Similarly, for $a=0$, Bardeen black holes with $g= 0.47165M$ and $g= 0.64025M$, respectively, lead to $\delta=-0.04$ (consistent with the Keck) and $\delta=-0.08$ (consistent with the VLTI). The constrained parameter space is shown in Figs.~\ref{fig:BarAng} and \ref{fig:BarDelta}. Clearly, rotating Bardeen black holes and no-horizon spacetime, within a finite parameter space, categorically satisfy the observed Sgr A* black hole shadow observables. Even though the EHT did not give any bound on the Sgr A* black hole shadow $\Delta C$, we show the circularity deviation $\Delta C$ as a function of ($a,g$) in Fig.~\ref{fig:BarAsymm}. Evidently, for black hole shadows, $\Delta C\leq 0.04$ and no-horizon spacetime shadows, $\Delta C\leq 0.017$.

Figures \ref{fig:HayAng}, \ref{fig:HayDelta}, and \ref{fig:HayAsymm}, respectively, show the Hayward black holes along with the corresponding no-horizon shadow observables $\theta_{sh}$, $\delta$ and $\Delta C$. We find that, for $a=0$, Hayward black holes with $g \geq  0.5302M$ and no-horizon spacetimes with $ g \leq 1.11921M$ cast shadow radii with $\theta_{sh}\in (46.9, 50)\mu$as corresponding to the observed Sgr A* shadow angular diameter (cf. Fig.~\ref{fig:HayAng}). Similarly, for $a=0$, Hayward black holes with $g= 0.9768M$ and no-horizon spacetime with $g= 1.16356M$, respectively, leads $\delta=-0.04$ (consistent with the Keck) and $\delta=-0.08$ (consistent with the VLTI) (cf. Fig.~\ref{fig:HayDelta}). The Hayward no-horizon spacetime shadows lead to the maximum $\delta=-0.103376$ for $a=0$ and $g=1.2183407M$. It is interesting to note that both the Hayward black holes and the no-horizon spacetime are consistent models for the Sgr A* shadow. Moreover, Hayward black hole shadows are bounded by $\Delta C\leq 0.06$, whereas no-horizon spacetime shadows $\Delta C\leq 0.0304$. 

Figures \ref{fig:NSAng}, \ref{fig:NSDelta}, and \ref{fig:NSAsymm}, respectively, show the Ghosh-Culetu black holes and corresponding no-horizon shadow observables $\theta_{sh}$, $\delta$ and $\Delta C$ in the parameter space ($a, g$). It is important to note that the Ghosh-Culetu black hole shadows cover a large window $\theta_{sh}\in (33.72, 50.28)\mu$as.  For $a=0$, Ghosh-Culetu black holes with $0.18345M\leq g \leq  0.62058M$ cast shadow radius with $\theta_{sh}\in (46.9, 50)\mu$as corresponding to the observed Sgr A* shadow angular diameter, whereas the corresponding no-horizon spacetimes shadow size fall outside the $1\sigma$ bound $41.7-55.6\,\mu$as. In particular, the no-horizon spacetime with $a=0$ and $g=1.26453M$ produced the smallest shadow with an angular diameter of $\theta_{sh}=30.6156\mu$as. Similarly, for $a=0$, Ghosh-Culetu black holes with $g= 0.483224M$ and $g= 0.673512M$, respectively, lead to $\delta=-0.04$ (Keck) and $\delta=-0.08$ (VLTI). In contrast to Bardeen and Hayward black holes, the Ghosh-Culetu black holes parameters are tightly constrained from the Sgr A* shadow (cf. Figs.~\ref{fig:NSAng} and \ref{fig:NSDelta}).  As a result, the Sgr A* black hole shadow is consistent with the Ghosh-Culetu black holes and not with the corresponding no-horizon spacetimes. Recently, constraints on the Ghosh-Culetu black hole parameters have been derived from the angular diameter of the shadow of the Sgr A* black hole \citep{Banerjee:2022iok}.

We calculated and depicted the Simpson-Visser black hole and corresponding no-horizon spacetime shadow angular diameter $\theta_{sh}$ in Fig.~\ref{fig:SVAng}, the shadow angular diameter deviation $\delta$ in Fig.~\ref{fig:SVDelta}, and the circularity deviation $\Delta C$ in Fig.~\ref{fig:SVAsymm}. It is worthwhile to mention that the Simpson-Visser spacetime shadows are completely independent of the parameter $g$. The rotating Simpson-Visser black hole and no-horizon spacetimes cast shadows with an angular diameter $\theta_{sh} \in (47.4074, 50.2831)\mu$as. In particular, Simpson-Visser black holes with $a \geq  0.3722M$ cast shadow radius with $\theta_{sh}\in (46.9, 50)\mu$as corresponding to the observed Sgr A* shadow angular diameter. Similarly, Simpson-Visser black holes with $a = 0.8884M$ cast a shadow that leads to $\delta=-0.04$ and satisfies the Keck bound for Sgr A* black holes.

\section{Conclusion}\label{sect5}
The supermassive black hole, Sgr A*, at the center of the Milky Way, is an ideal and realistic laboratory for testing the strong-field predictions of general relativity. Recently, in a series of papers, the EHT collaboration released the first horizon-scale images of Sgr A*, acquired at a wavelength of 1.3 mm \citep{EventHorizonTelescope:2022exc,EventHorizonTelescope:2022urf,EventHorizonTelescope:2022vjs,EventHorizonTelescope:2022wok,EventHorizonTelescope:2022xnr,EventHorizonTelescope:2022xqj}. A bright ring of emission that enclosed a central brightness depression characterizes the images \citep{EventHorizonTelescope:2022wok}. 
The EHT collaboration's predicted size of shadow is based on the prior information on the mass-to-distance ratio of the Sgr A* black hole. The EHT results are consistent with the prediction on the Kerr metric, and there is no evidence for any breaches of the theory of GR.
Here, we used the EHT observation of the black hole shadow in Sgr A* to place constraints on deviation parameters associated with four well-motivated (non-Kerr) rotating regular black holes.

The rotating regular spacetimes considered here possess a de Sitter core, except for the Ghosh--Culetu spacetime, which has a Minkowski core and a Simpson-- Visser spacetime with a bounce geometry. The latter is of particular interest because, depending on the value of the regularization parameter $g$, it can describe either a regular black hole or a traversable wormhole. The shadows of Bardeen, Hayward, and Ghosh-Culetu rotating regular black holes are smaller and more distorted than those of the Kerr black hole. On the other hand, the Simpson-Visser black hole shadow is identical to that of a Kerr black hole. We have found that, for a finite parameter space, these rotating regular spacetimes, even in the horizon's absence, lead to a closed photon ring. The Simpson-Visser no-horizon spacetime with an arbitrarily large spin can cast a closed photon ring, whereas Bardeen, Hayward, and Ghosh-Culetu no-horizon spacetimes are restricted to small values of $a$, $a \lesssim 0.5M$. We have further charcaterized the shadows using the angular diameter $\theta_{sh}$, the shadow angular diameter deviation from the Schwarzschild black hole shadow $\delta$, and the circularity deviation $\Delta C$. Using the bounds on $\theta_{sh}$ and $\delta$ for the Sgr A*, we derive constraints on the spacetime parameters ($a, g$).

The shadows of Bardeen black holes with $(a=0,\;\; 0.18M\leq g \leq  0.595M)$ and $(a=0.2M,\;\; 0.154M\leq g\leq 0.58367M)$ are consistent with the Sgr A* black hole shadow angular diameter $\theta_{sh}$, whereas, the corresponding no-horizon spacetime with $g \leq 0.83794M$ also satisfies the $1\sigma$ bound. In comparison, the Sgr A* shadow angular size placed a relatively weaker constraint on Hayward spacetime, i.e.., black holes with $g \geq  0.5302M$ and no-horizon spacetimes with $ g \leq 1.11921M$ produced a shadow that is consistent with the Sgr A* shadow size. On the other hand, the Ghosh-Culetu spacetime parameters are tightly constrained, e.g., for $a=0$, $0.18345M\leq g \leq  0.62058M$ and for $a=0.20M$, $0.155M\leq g\leq 0.61116M$. While the photon rings of Bardeen, Hayward, and Simpson-Visser no-horizon spacetime are consistent with Sgr A* shadow size, the photon ring of Ghosh-Culetu no-horizon spacetime is considerably smaller and falls outside the $1\sigma$ bound. For mathematical calculations, we have considered the inclination angle $\theta_o=50^o$, as relevant for the Sgr A* black hole. If we consider different inclination angle, the Sgr A* shadow observables will lead to weaker/stronger constraints on black hole parameters.

The Sgr A* shadow fits with the projections of the Kerr metric, which, according to the GR summarized no-hair theorem, represents any astrophysical black holes spacetimes. The celebrated theorem may not hold for the modified theories of gravity that admit non-Kerr black holes. Our study indicates that rotating regular spacetimes, except possibly Ghosh-Culetu no-horizon spacetime, agree with the results of SgrA* and they are strong candidates for the astrophysical black holes. 

Given the significance of the accretion onto black holes, relativistic magnetohydrodynamic simulations of the hot plasma flow with the effects of magnetic fields in the accretion disk are necessary for a proper analysis of shadow. 
One can, however, take the existing results as an encouraging sign to analyze these effects.

\appendix
\section{Regularity of the rotating metric}\label{sec:appendix}

For the rotating metric (\ref{rotmetric}), the metric function diverges for $\Sigma=0$, which is $r=0,\theta=\pi/2$. However, for the mass functions discussed in subsections \ref{sec:bar}--\ref{sec:sv}, the rotating metric (\ref{rotmetric}) defines a globally regular spacetime. This can be seen from the expressions of its curvature invariant, namely Ricci scalar, Ricci square, and Kretschmann scalar  \citep{Bambi:2013ufa}. We calculate these curvature scalars for all four of the rotating regular black holes. For the Bardeen and the Hayward spacetimes, they reduce to simpler forms at the origin $(r=0,\theta=\pi/2)$ 
\begin{eqnarray}
\lim_{r\to 0}\Big( \lim_{\theta\to \pi/2} R\Big)&=&- \frac{24 M}{g^3},\nonumber\\
\lim_{r\to 0}\Big( \lim_{\theta\to \pi/2} R_{\mu\nu} R^{\mu\nu}\Big)&=&\frac{144 M^2}{g^6},\nonumber\\
\lim_{r\to 0}\Big( \lim_{\theta\to \pi/2} R_{\mu\nu\alpha\beta}R^{\mu\nu\alpha\beta}\Big)&=& \frac{96 M^2}{g^6}.\nonumber
\end{eqnarray}
Clearly, the curvature scalars are finite at the black hole center ring as long as $g\neq 0$. Whereas for the Ghosh-Culetu spacetime, all the curvature scalars vanish at $(r=0,\theta=\pi/2)$ \citep{Ghosh:2014pba}:
\begin{eqnarray}
\lim_{r\to 0}\Big( \lim_{\theta\to \pi/2} R\Big)=
\lim_{r\to 0}\Big( \lim_{\theta\to \pi/2} R_{\mu\nu} R^{\mu\nu}\Big)=
\lim_{r\to 0}\Big( \lim_{\theta\to \pi/2} R_{\mu\nu\alpha\beta}R^{\mu\nu\alpha\beta}\Big)=0.\nonumber
\end{eqnarray}
In contrast, for the Simpson-Visser spacetime the surface $r = 0$ is an oblate spheroid of size $g$. Only when $g=0$, the spheroid collapse to a ring at $\theta=\pi/2$ and the usual singularity of the Kerr geometry isthen  recovered. When instead $g \neq 0$, the singularity is excised and $r = 0$ is a regular surface of finite size. The curvature scalars read
\begin{eqnarray}
\lim_{r\to 0}\Big( \lim_{\theta\to \pi/2} R\Big)&=&-\frac{2 \left(2 a^4+a^2 g (g-6 M)+g^2 (g-2 M)^2\right)}{a^2 g^4},\nonumber\\
\lim_{r\to 0}\Big( \lim_{\theta\to \pi/2} R_{\mu\nu} R^{\mu\nu}\Big)&=&\frac{2 \left(4 a^8+4 a^6 g (g-4 M)+a^4 g^2 \left(4 g^2-19 g M+34 M^2\right)+a^2 g^3 (g-2 M)^2 (2 g-9 M)+g^4 (g-2 M)^4\right)}{a^4 g^8},\nonumber\\
\lim_{r\to 0}\Big( \lim_{\theta\to \pi/2} R_{\mu\nu\alpha\beta}R^{\mu\nu\alpha\beta}\Big)&=& \frac{4 \left(4 a^8+2 a^6 g (2 g-3 M)+a^4 g^2 \left(3 g^2-8 g M+12 M^2\right)+2 a^2 g^3 (g-2 M)^2 (g-3 M)+g^4 (g-2 M)^4\right)}{a^4 g^8}.\nonumber
\end{eqnarray}
Here, $R$, $R_{\mu\nu}$ and $R_{\mu\nu\alpha\beta}$ are, respectively, the Ricci scalar, Ricci tensor, and Riemann tensor. 

\section{Acknowledgement}
S.G.G. would like to thank SERB-DST for project No. CRG/2021/005771. R.K.W. would like to thank the University of KwaZulu-Natal for the continued support and the NRF for the postdoctoral fellowship. S.D.M acknowledges that this work is based upon research supported by the South African Research Chair Initiative of the Department of Science and Technology and the National Research Foundation.



\bibliography{ChiSqr}
\bibliographystyle{aasjournal}
\end{document}